# Uncovering the deformation mechanism of glasses during indentation through high-resolution X-ray scattering

*M. Faizal Ussama Jalaludeen*[1], *Søren S. Sørensen*[1], *Johan F. S. Christensen*[1], *Anders K. R. Christensen*[1], *Sidsel Mulvad Johansen*[1], *Samraj Mollick*[1], *Yuanzheng Yue*[1], *Sharafat Ali*[2], *Sebastian Kalbfleisch*[3], *Morten M. Smedskjaer*[1,*]

[1]Department of Chemistry and Bioscience, Aalborg University, DK-9220 Aalborg, Denmark

[2]Department of Built Environment and Energy Technology, Linnæus University, SE-351 95 Växjö, Sweden

[3]MAX IV Laboratory, Lund University, PO Box 118, S-221 00 Lund, Sweden

*Corresponding Author. Email: mos@bio.aau.dk

## Abstract

Indentation experiments can be used to mimic real-life damage events of glasses that lead to surface flaws and thus lower practical strength. Conventional indentation studies often focus on the surface deformation after unloading. However, to understand the link between the surface deformation and structure, it is crucial to characterize the sub-surface deformation during the indentation process. The indentation-induced deformation, consisting of both elastic and plastic zones, is governed by the glass composition and structure, indentation and atmospheric conditions, and stress state. However, only a few experimental methods exist for characterizing the sub-surface indentation deformation mechanism during indentation. In this study, we use synchrotron X-ray nanoscattering to probe the deformation mechanism *in situ* during indentation of four types of oxide and oxynitride glasses with distinct structural features. This is done by measuring the variation in the position and intensity of the first sharp diffraction peak of the X-ray structure factor with a high spatial resolution down to ~100 nm. We find that the deformation zones of these glasses, which are characterized by the shape, size, and relative contribution between densification and shear flow under different indentation loads, vary with Poisson's ratio. Thus, our work provides new insights into the mechanical behavior of oxide glasses, contributing to the design of more damage-resistant glasses.

## 1. Introduction

Oxide glasses remain as the commercially most important glass family owing to their high optical transparency, relatively high hardness, excellent workability and low cost of raw materials and processing [1]. However, they suffer from high brittleness and catastrophic failure upon the application of excessive stress, mainly due to stress-concentration at surface defects (flaws) and the lack of energy dissipative deformation mechanisms. In turn, this makes their practical strength orders of magnitude smaller than their theoretical strength [2, 3]. While being macroscopically brittle, oxide glasses can undergo local plastic deformation during sharp contact (indentation) loading [4]. Mechanisms that lead to such plastic deformation are generally classified as densification and volume conservative shear (viscous) flow, with all glasses generally featuring a combination of both mechanisms [5]. However, the underlying micro-mechanical origins that give rise to these types of deformation remain largely unknown. Revealing these origins is important for understanding the intricate relation between deformation, stress release, and microscopic crack initiation.

It has been proposed that the resistance to crack initiation (i.e., the critical indentation load required to initiate cracking) is positively correlated to the densification ability of the glass under the indenter tip [6], a property that is highly challenging to measure *in situ* [7]. In turn, the indentation-induced densification in oxide glasses depends on their Poisson's ratio [8]. The typical method used to study indentation deformation mechanism involves surface topography measurements (e.g., using atomic force microscopy) of indented samples before and after a heat-treatment to quantify the relative contributions of densification and shear flow through a measured volume recovery [8]. However, only a part of the indentation-induced damage is observed on the surface, with a major part concealed beneath the surface. Analysis of this sub-surface region is highly challenging and often requires destructive preparation techniques of the glass. Possible techniques to analyse the densification and damage zones include chemical dissolution methods [9], cross-sectional investigation by focused ion-beam milling often coupled with scanning electron microscopy [10, 11], micro-Raman spectroscopy [12-14], and micro-focused Brillouin light scattering [15, 16].

These measurements are performed *ex situ* after unloading and do not provide nanoscale resolution of the densification zone; therefore, only limited spatiotemporal information is obtained. This is important to address, since it is known from both molecular dynamics simulations [17] and scattering experiments [18, 19] that the deformation mechanism can be reversible upon unloading, with structural changes occurring at the atomic scale. In addition, some techniques have the risk of distorting the stress field around the area being removed, e.g., upon etching or formation of the cross-section. Yoshida et al. [8] have proposed an indirect quantification of the indentation-induced sub-surface densification in various oxide glasses by post-annealing volume recovery measurements of the indented regions, but this method lacks temporal information. Recent studies have attempted to map the atomic glass structure with microscale resolution using non-destructive *in situ* micro-Raman spectroscopy during indentation [20, 21], but it has remained challenging to visualize the deformation zone during the indentation process at sufficiently high resolution. *In situ* spatiotemporal mapping of the glass sub-surface during indentation is thus necessary to get the true picture of the deformation process [22]. We note that photo-elastic measurements based on birefringence imaging capture the elastic deformation response of glasses *in situ* (during indentation) and the elasto-plastic deformation response *ex situ* (after unloading), enabling quantitative evaluation of non-linear mechanical behaviour [23]. However, this method does not probe the complete deformation zone during the indentation process. In contrast, synchrotron-based X-ray techniques offer an attractive alternative given their possibility for nanoscale resolution, real-time non-destructive analysis, and flexibility in terms of sample environments [24-26].

As shown in earlier works [18, 19, 26], *in situ* investigation of glasses during indentation by X-ray nano-scattering can indeed be used to map the local glass deformation over the probed volume. This is made possible by using a nano-sized X-ray beam, which provides local structural information from a small region of the glass being probed, from which densification and other types of deformation (e.g., strain) can be extracted. Raster scanning over an area is done by moving the samples together with the indenter while collecting the X-ray scattering signals from the sample being probed. These signals can then be compared with one another to form a spatial 2D map to understand the changes in the glass structure due to an induced external stress from nano-indentation. In detail, the local X-ray scattering pattern [18, 19, 26, 27] can be obtained with sufficient inverse-space ($q$ – scattering vector) resolution to resolve the first sharp diffraction peak (FSDP), which is a typical signature of the medium-range order (MRO) structure of oxide glasses. Assignments of the FSDP include local tetrahedral ordering [28], random packing of the structural units [29], pseudo or quasi Bragg reflection from disordered planes or rings [30], scattering by clusters of definite size [31, 32], density fluctuations [33], structural connectivity up to medium range orders [34], or the shape of the rings in the glass network [35]. As such, the position and shape of the FSDP are sensitive to the glass composition and processing history conditions [31]. While the exact physical mechanisms behind the FSDP of glasses is still being debated, changes in FSDP upon indentation can be used a metric for estimating the structural deformation endured by glasses.

Building upon our recently established methodology [18], we here probe the local variation in the position of the FSDP during indentation of four different glass compositions. Compared to the previous work, we achieve this with a very high spatial resolution (~100 nm) based on the X-ray optics and high flux available at the NanoMAX beamline of the MAX IV Laboratory [36]. Through appropriate nanoindentation conditions, we induce crack-free indents (although cracking often occurs during unloading) and thus map the deformation zone *in situ* (i.e., unaffected by cracking). We compare the deformation response at multiple loads between oxide and oxynitride glasses (silica, sodium-calcium silicate, sodium aluminosilicate, and oxynitride calcium aluminosilicate) with varying Poisson's ratio and thus mechanical response in terms of the extent of densification to shear flow [37]. Since Poisson's ratio directly reflects the structural properties of the glass network, it provides an estimate of the magnitude of densification and consequently the indentation deformation mechanism [38]. In this work, we also propose a method for quantifying the deformation zone size that does not require post-indentation annealing of glasses, thereby preserving the glass' intrinsic deformation state. Using this method, we decouple the elastic from plastic contributions to the deformation zone. Based on the deformation data derived from this nano-scale scattering analysis, we then correlate the dynamically changing (indentation-induced) glass structure with the resulting mechanical behaviour. Overall, our results contribute to identifying and understanding the micro-mechanical features controlling the damage resistance of oxide and oxynitride glasses.

## 2. Experimental procedure

*Sample preparation.* The compositions and basic properties of the four glasses studied in this work are given in Table 1. The vitreous silica (v-$SiO_2$) used in this study was a commercial High Purity Fused Silica of 7980 Standard Grade from Corning Incorporated [39], with low metallic impurities (<1 ppm) and a hydroxyl concentration of 800-1000 ppm. The sodium calcium silicate (NCS-80) and sodium aluminosilicate (NAS-60) glasses were prepared using melt-quenching in an electrical furnace (SF17, Entech, Ängelholm, Sweden). The raw materials, $SiO_2$ (Sigma-Aldrich, >99.8%), $CaCO_3$ (Chemsolute, >99.5%), $Na_2CO_3$ (Honeywell International, ≥99.5%), $Al_2O_3$ (Sigma-Aldrich, >99.5 %) were melted in a Pt-Rh crucible, homogenized for 4 hours in the temperature range of 1550 and 1650 °C, before being quenched onto a brass plate. The melt-quenched glasses were then immediately transferred to a preheated muffle furnace and annealed at their estimated glass transition temperature ($T_g$) to relieve the internal stresses. Details on the preparation of the oxynitride calcium aluminosilicate (CAS-N) glass is given elsewhere [40].

**Table 1.** Composition, density ($\rho$), Poisson's ratio ($v$), glass transition temperature ($T_g$), linear X-ray attenuation depth ($1/\mu$) at 14 keV, and final thickness ($t$) of the studied glasses.

| Sample | $SiO_2$ (mol%) | $Na_2O$ (mol%) | CaO (mol%) | $Al_2O_3$ (mol%) | AlN (mol%) | O/N (-) | $\rho$ (g/cm³) | $v$ (-) | $T_g$ (°C) | $1/\mu$ (μm) | $t$ (μm) |
|---|---|---|---|---|---|---|---|---|---|---|---|
| v-$SiO_2$ | ≈100 (OH: 800-1000 ppm) | - | - | - | - | - | 2.198 ± 0.002 | 0.166 | 1017 | 640 | 136.8 ± 1.2 |
| NCS-80 | 81.00 | 9.54 | 9.46 | - | - | - | 2.454 ± 0.001 | 0.203 | 582 | 932 | 136.6 ± 2.1 |
| NAS-60 | 60.62 | 20.59 | - | 18.80 | - | - | 2.415 ± 0.001 | 0.218 | 731 | 1706 | 130.0 ± 2.1 |
| CAS-N | 32.86 | - | 31.52 | 15.73 | 19.89 | 7.26 | ×2.840 ± 0.010 | 0.278 | 890 | 1477 | 137.6 ± 0.6 |

*Thermal and physical properties.* Differential scanning calorimetry (DSC) measurements (Netzsch STA 449F1) were carried out on solid glass pieces with proper baseline corrections to determine their $T_g$ values. After determining the $T_g$ (see Supporting Figure S1), the glasses were re-annealed at their respective $T_g$ for 30 minutes. The actual compositions of the glasses were determined using inductively coupled plasma optical emission spectroscopy, while density ($\rho_{glass}$) was determined using Archimedes' principle of buoyancy with absolute ethanol as a reference medium / immersion medium ($\rho_{reference}$),

$$\rho_{glass} = \rho_{reference}\left(\frac{D}{D - S}\right) \tag{1}$$

Here, $D$ and $S$ are the dry and suspended weight of the glass sample, respectively.

Poisson's ratio ($\nu$) of the glasses was calculated using ultrasonic pulse-echo overlap technique from the measured longitudinal and transverse waves, as

$$\nu = \frac{(V_L^2 - 2V_S^2)}{[2(V_L^2 - V_S^2)]} \tag{2}$$

where $V_L$ is the longitudinal wave and $V_S$ is the transverse wave.

*Sample thinning and thickness optimization.* For the indentation and X-ray scattering experiments, we used initial glass samples of approximately 5-10×5×5 mm³ that were ground and polished to an optical finish on one of the faces. The glass samples were then reduced to a thickness of 130 to 200 μm on the sides before being polished to optical finish on both ends (entry and exit) for the X-ray scattering and indentation measurements. The first steps of grinding and polishing were done in absolute ethanol using 220 to 4000 grit SiC papers, while the final steps were done using water-free diamond suspensions (3 and 1 µm) to obtain optical finish and avoid surface hydration. Theoretical interaction (attenuation) of different glass samples of certain thickness as a function of the X-ray energy is shown in Supporting Figure S2 [41]. The calculation of the linear attenuation coefficient (see values in Table 1) is based on Beer-Lambert's equation for the linear attenuation of a homogeneous material, considering the measured sample compositions and bulk densities.

The sample thicknesses (135±5 µm, see Table 1) were then chosen based on data from preliminary optimization experiments on glasses showing similar X-ray scattering cross-section as well as from the X-ray interaction cross-section calculations for glasses of varying thicknesses (Supporting Figure S3). Another reason for not exceeding the sample thickness beyond 200 µm was the limiting depth of focus of the X-rays (234 µm, based on the beam operating conditions given below). These thin glass samples were mounted on a sample holder (Ted Pella Inca) with a thermopolymer glue (Crystalbond™ 509). A 136° face-angled diamond wedge indenter tip with a length of ~205 µm was custom-made by Synton-MDP (Switzerland). The thickness of the wedge indenter tip was intentionally made to be larger than the sample thickness to minimize frictional effects during contact loading. However, the use of a longer indenter tip cast a shadow and interfere with part of the scattering signal coming from the sample.

*X-ray nanoscattering.* The X-ray nanoscattering measurements were performed *in situ* using a nanoindenter sample environment [42] (see details below) at the NanoMAX beamline of MAX IV Laboratory. Information on the beam optics is given elsewhere [43]. The experiments were conducted under ambient conditions. An X-ray beam energy of 14 keV ($\lambda \sim 0.8856$ Å) was operated in high flux mode yielding $3 \cdot 10^{10}$ photons $s^{-1}$ [44] with an exposure time of 50 ms per point on the sample. This provided interaction cross-sections (Supporting Figure S2) to ensure sufficient statistics to resolve the FSDP, with a beam spot size of radius 100 nm.

The scattered signals were captured using a Pilatus3 1M detector placed at a distance of 150 mm away from the sample in transmission geometry. An example of the scattering signal obtained with the detector is shown in Supporting Figure S4. This setup provided a $q_{max}$ of around 5 Å$^{-1}$ when the beam was centred on the detector. Beam calibration was done using $LaB_6$ crystalline powder and dead pixels in the detector were masked prior to measurements on the sample (Supporting Figure S5). The setup included a scatter shield that restricted and suppressed stray air radiation from hitting the sample at the incident end, which would otherwise have caused secondary scattering effects, adding background noise. Furthermore, a He gas filled flight tube was placed in between the sample and detector to avoid scattering loss by air absorption or the secondary effects from scattered X-rays [44]. The flight tube had an uncoated 2.5 µm thick Mylar film on the side facing the sample, while the exit window side was made of aluminized Mylar (i.e., an 8 µm thick polyester film coated with 50 nm of Al), which produced additional scattering as highlighted in Supporting Figure S6. However, this additional scattering from Mylar is outside our $q$ region of interest and is of very low intensity. The measured total scattering intensity $I(q)$ was obtained for grid points in an area of ~ 25×25 µm$^2$ over 250 points per line on the sample by raster scanning at each holding load step as well as before and after indentation loading, as detailed in the following.

*In situ indentation.* The indenter setup was placed within the X-ray beam as shown in Figure 1a, along with various other components. This included the top and on-axis microscopes that enabled sample alignment with respect to the indenter, as well as the entire indenter in line with the X-ray beam. Scatter shield, flight tube, and the various motors that move different components independently are also visible. The indenter setup housed the load cell, indenter tip driven by piezo actuator, and the sample holder (Figure 1b).

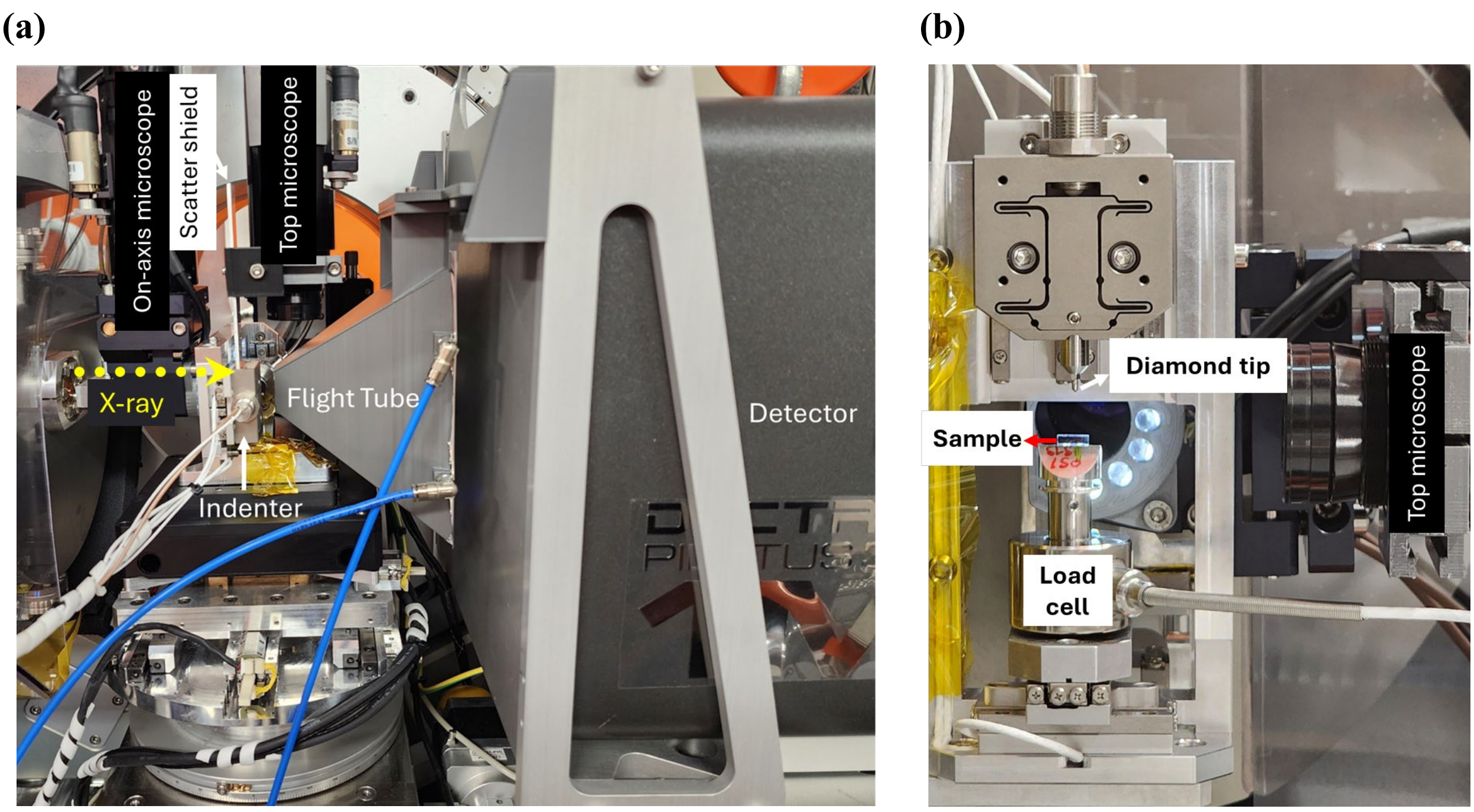

**Figure 1.** (a) Side view of the experimental setup, comprising the components used for alignment of the sample with the X-ray beam (coming from the narrow gap through the optics on the left, see yellow arrow), optical microscopes (placed on-axis with the X-ray beam and perpendicular to the beam), and components for suppressing the stray and air scattering before and after the sample, respectively. (b) Front view of the standard nanoindenter setup, housing the glass sample and the 136° diamond wedge indenter tip.

The indentation experiments were performed under quasi-static conditions. This was done using a standard nano-indenter equipped with 2.5 N load cell (Alemnis AG). A maximum load of 2 N was applied on all glasses, with an intermediate holding step at 1 N. This was chosen considering the available experiment time, as well as to avoid cracking of the glasses during indentation. As mentioned above, we used a customized 200-µm diamond wedge indenter tip and face angle of 136° (i.e., same face angle as a four-sided pyramid Vickers tip) to indent the glasses. The measurements were initiated using the Alemnis micro-indenter control software (AMICS) that collected the load-displacement data. The regions in and around the indentation site were then raster scanned using the X-ray beam as linear motors moved the indenter setup along with the sample with respect to the beam. As cracking in glasses is affected by atmospheric conditions, humidity and temperature data were collected over the experimental period. The mean temperature and relative humidity were measured to be 24.8±0.7 °C and 37.7±1.2 %, respectively, over the course of all the indentation experiments.

*Beam damage scans. In situ* nanoscattering of indentation experiments require localized exposing specific regions of the glass sample to the intense and focused X-ray radiation, which may induce phenomena such as structural changes (including defect generation), local heating etc., [46]. Any such structural modification other than those caused by indentation makes it complicated to understand the 'true' deformation behavior of these glasses. To investigate this issue, the glass samples were therefore exposed multiple times (albeit without indentation) to mimic the *in situ* scanning sequence and the collected $I(q)$ data were then analyzed to check for any changes due to beam damage.

*Data analysis.* As the X-ray passes through the sample, they become absorbed, scattered, and transmitted before reaching the detector (see schematic illustration in Figure 2a). The magnitude of the scattering wavevector was given as,

$$q = \frac{4\pi \sin \theta}{\lambda}. \tag{3}$$

$q$ and the inter-planar spacing ($d$) are approximately related through,

$$q \approx \frac{2\pi}{d}. \tag{4}$$

An example of the raw detector image obtained for a frame is shown in Supporting Figure S5. The detector image was masked appropriately to remove shadows from the beam stop, indenter tip, and dead pixels. After masking, the scattering signal thus obtained per frame was integrated azimuthally (over the full range) to get the integrated intensity $I(q)$ curve, as shown in Figure 2b. The detector images were all processed and azimuthal integration was performed using azint algorithm [48] at the beamline. The integrated signal contains the FSDP, from which we extracted features such as the peak position ($q_{max}$ or $q_{FSDP}$), and intensity, which will be used for further 2D mapping.

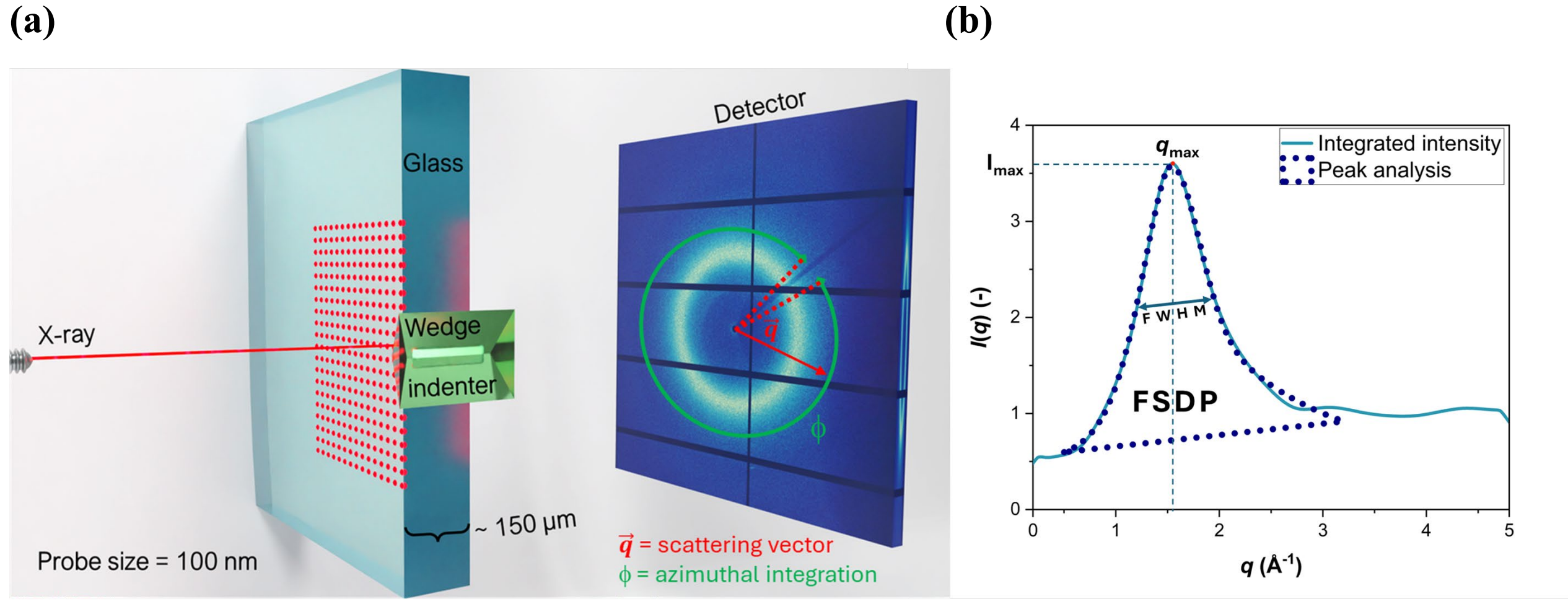

**Figure 2.** (a) Schematic of the *in situ* X-ray indentation experiment. (b) Example of integrated intensity data for which the peak analysis was performed.

*Data processing.* An example of the indentation (loading, holding, and unloading) sequence followed in this study is shown in Supporting Figure S7 for vitreous silica. The glass was indented in two steps up to 2 N before unloading the indenter. Once the glass was indented to a specific load, the load was held constant for approximately 3 min for the glass to stabilize it in terms of creep due to time-dependent deformation. After this creep stabilization, the displacement was held constant, before raster scanning the area around the indented area with X-rays. This also allowed the glass to be fixed in position with respect to the indenter during mapping.

## 3. Results and Discussion

*FSDP position maps:* Before indentation, the glass is scanned with X-rays over an area of ~25×25 µm$^2$ to obtain the scattering intensity $I(q)$ of the undeformed glass. As shown in Figure 3a, a 2D map is thus generated with horizontal and vertical coordinates corresponding to the spatial region on the sample being scanned and the color scale showing the change in FSDP position relative to that before applying an indentation load at the same location. The 2D map prior to indentation appears homogeneous (structurally uniform), as expected for an unstrained and compositionally homogeneous glass. We also note that all the 2D maps contain some portion of background air scattering data above the sample surface, which is collected to better identify the sample edge as well as to subtract the air scattering from the collected data of the sample. Hence, after proper background subtraction, black regions corresponding to $q_{max}$ values below $q_{FSDP}$ of the undeformed region prior to indentation originate from air scattering, which contributes a broad background signal over the measured $q$-range (Supporting Figure S6). This background is, however, too weak to shadow the signal from the sample. The absolute FSDP position maps are shown in Supporting Figure S8. Finally, we note that besides the effect of indentation, structural changes can occur due to beam-induced sample damage because of the repetitive probing of the X-ray beam over the same region on the sample. This is demonstrated from the overview scan (50×50 $\mu m^2$) shown in Supporting Figure S9), i.e., following the indentation-based scans (four scans), repeatedly exposed areas display altered FSDP features (intensity, $q$-position and shape) compared to previously unexposed regions (one scan) of the sample. This phenomenon (Supporting Text S1) is found to be dependent on glass composition, i.e., some glasses are more susceptible than others (Supporting Figures S9-S10). However, even though we thus observe beam-induced damage for some glass compositions, the effect of beam damage is corrected for by comparing the relative shift in FSDP with respect to that of the undeformed region.

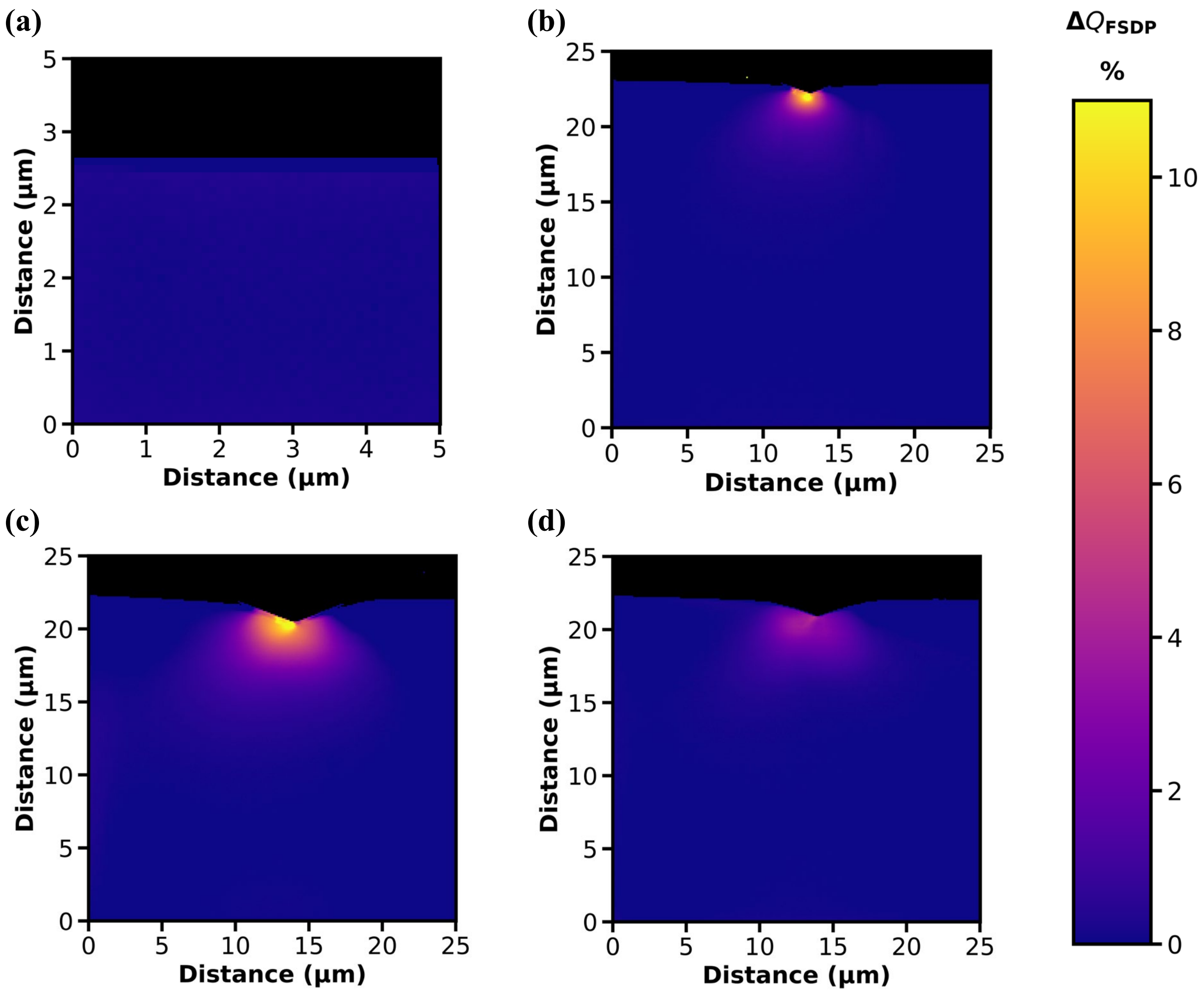

**Figure 3.** 2D maps showing the relative shift of the FSDP position as obtained from the $I(q)$ data for vitreous silica: (a) before indentation, (b) during indentation loading at 1 N, (c) during indentation loading at 2 N, and (d) after indentation unloading. The color bar on the right of the 2D maps indicate the percentage change shift in $q_{\mathrm{FSDP}}$ relative to the undeformed $q_{\mathrm{FSDP}}$ prior to indentation.

*Vitreous silica deformation.* Upon indenting vitreous silica at a load of 1 N, the FSDP shifts towards higher $q$ values around the tip of the indenter within the glass (Figure 3b). This increase in FSDP position towards higher $q$ corresponds to a decrease of the correlation distance (reduction of free volume) in the MRO glass structure, i.e., densification. The deformation zone becomes larger with increasing load (from 1 to 2 N, see Figure 3c). Upon unloading, the pure elastic deformation recovers, leaving a zone of permanent plastic deformation (Figure 3d). This trend of FSDP shifting to higher $q$ and the increasing size of the deformation zone with increasing indentation load is similar to our previous study on vitreous silica [18], but herein it is obtained at significantly higher spatial resolution.

To isolate the deformation zone from the region unaffected by the indentation field, we employ a thresholding-based approach. In detail, we use the peak position information from the generated 2D position maps and generate histograms to understand the statistical nature of the deformation, including skewness, extremeness etc. An example of this is shown in Figure 4a. As the mapping was performed on a large area (25×25 μm$^2$) relatively to the size of the indented region, most of the data points in the heatmap are collected from regions unaffected by indentation. We aggregate these data points as Mode (most common data point in the distribution) in Figure 4a. Mode-based threshold is then applied to mark the boundary of the undeformed and deformed portions of the sample. The data points beyond the threshold are masked (see Figure 4b), allowing us to clearly segment the heatmap into regions with similar $q_{max}$ values. Finally, this also allows us to identify the limits of the deformation zone as highlighted in green in Figure 4b.

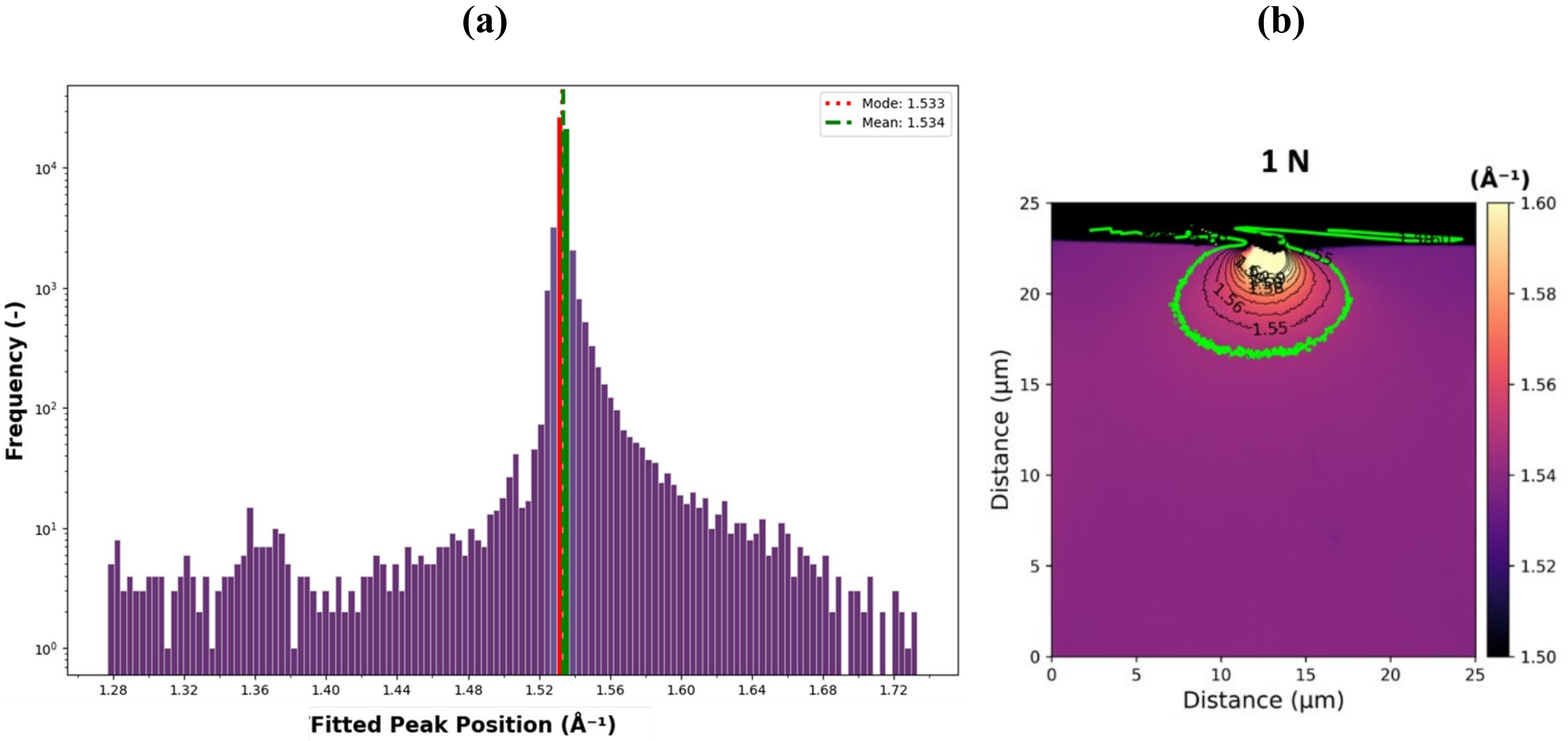

**Figure 4.** (a) Example of histogram of FSDP position values derived from the 2D heatmap including the background air scattering for vitreous silica at 1 N load. (b) Heatmap segmented using iso-lines of FSDP

position values, highlighting smaller subset regions within the deformation zone and the gradual change from maximum to minimum distortion (in green line) away from the point contact.

Based on this analysis, we then draw the boundary for the deformation zone, enabling us to further determine the size of the deformation zone by integrating the area within the limits. The resulting contour plot (Fig. 4b) not only segregates the deformation zone from the undeformed region but also provides a visualization of the variation within the deformation zone. For the case of vitreous silica, the distribution of FSDP position values appears unimodal and symmetric during loading. Also, as reported elsewhere [18], shear flow induced in vitreous silica by indentation only causes a change in FSDP intensity without a notable shift in FSDP position ($q_{FSDP}$). This deformation zone quantification thus assumes that either shear flow does not alter the mean FSDP position relative to the undeformed glass, or the FSDP shift reflects purely densification, in which case the densification zone is overestimated due to contributions from shear flow.

*Composition dependent deformation.* Having established the methodology for vitreous silica, we next compare the indentation-induced deformation of the four studied glasses. First, Supporting Figure S11a shows a comparison of the load-displacement response. The total work done during an indentation process is reflected in the area of the load-displacement curve, which denotes the amount of plastic deformation , and is found to decrease with increasing Poisson's ratio of glasses . As the glass is kept at a constant displacement during the scanning segment of our experiment, the glass undergoes structural relaxation (elastically-densified structures), which is reflected in the drop of load in Supporting Figure S11a. To this end, we calculate the relaxation or creep ratio $(P(t_0) - P(t))/P(t_0)$, where $P(t_0)$ is the starting load at which the displacement is held constant (1 or 2 N), and $P(t)$ is the load by the end of the displacement holding segment. A higher value of this ratio indicates a greater propensity for densification. As shown in Supporting Figures S11b-c, we indeed find that the ratio follows the densifying capability of the glass (i.e., higher for a highly densifying glass such as silica and lower for a less densifying such as oxynitride glass). The load drops at constant displacement could then be considered as relaxation of the densified structures. For glasses showing so-called "normal" indentation response, the penetration depth at a specific load follows an inverse relation with the glass hardness. The maximum displacement ($d_{max}$), except for silica, is lower for low Poisson's ratio glasses. After complete unloading, a permanent impression remains, which is determined as the final penetration depth ($h_f$). The ratio of this depth with respect to the total penetration depth ($h_{max}$) at the maximum load gives the permanently unrecovered deformation. This ratio $h_f/h_{max}$, as plotted in Supporting Figure S11d, decreases with increasing Poisson's ratio and is thus highest for vitreous silica (0.25).

As shown in Figure 3, the indentation deformation zone shape evolves with varying indentation loading conditions in the case of vitreous silica. During loading, the deformation zone is semi-circular, radiating outwards from a point contact and into the glass, and then becomes dumbbell-shaped after unloading, showing highly densified regions on the lateral sides of the point contact with respect to the indenter tip. This deformation zone shape matches that of a previous study done on vitreous silica [18]. The deformation maps for the NCS, NAS, and oxynitride glasses are shown in Figure 5. We find that this deformation zone remains semi-spherical in shape at all loading stages for NCS and NAS. For the high Poisson's ratio oxynitride glass, the zone has a V-groove shape, with the deformation being mostly concentrated close to the contact area (edge and faces) of the indenter tip, instead of extending axially inside the glass. As seen from Figure 5j,k and Supporting Figure S8d, the deformation zone of the oxynitride glass also consists of a zone of shear flow on either side of the densification zone at 1 N load, and a region of possible shear banding beneath the densification zone at 2 N. After the removal of load, the zone of shear banding disappears, leaving behind the permanent densification. As seen from Supporting Figure S8d, the shear flow region deforms elastically as it recovers upon unloading. Only the region directly beneath the indenter tip is permanently densified as seen from the shift in FSDP position maps. The shear flow region thus has higher $q_{FSDP}$ values due to elastic strain or densification during the loading cycle and the shift in $q_{FSDP}$ recovers back to similar peak position as of the undeformed glass region after unloading. We find that under any applied load, the morphology (shape and size) of the indentation deformation zone differs between the glasses.

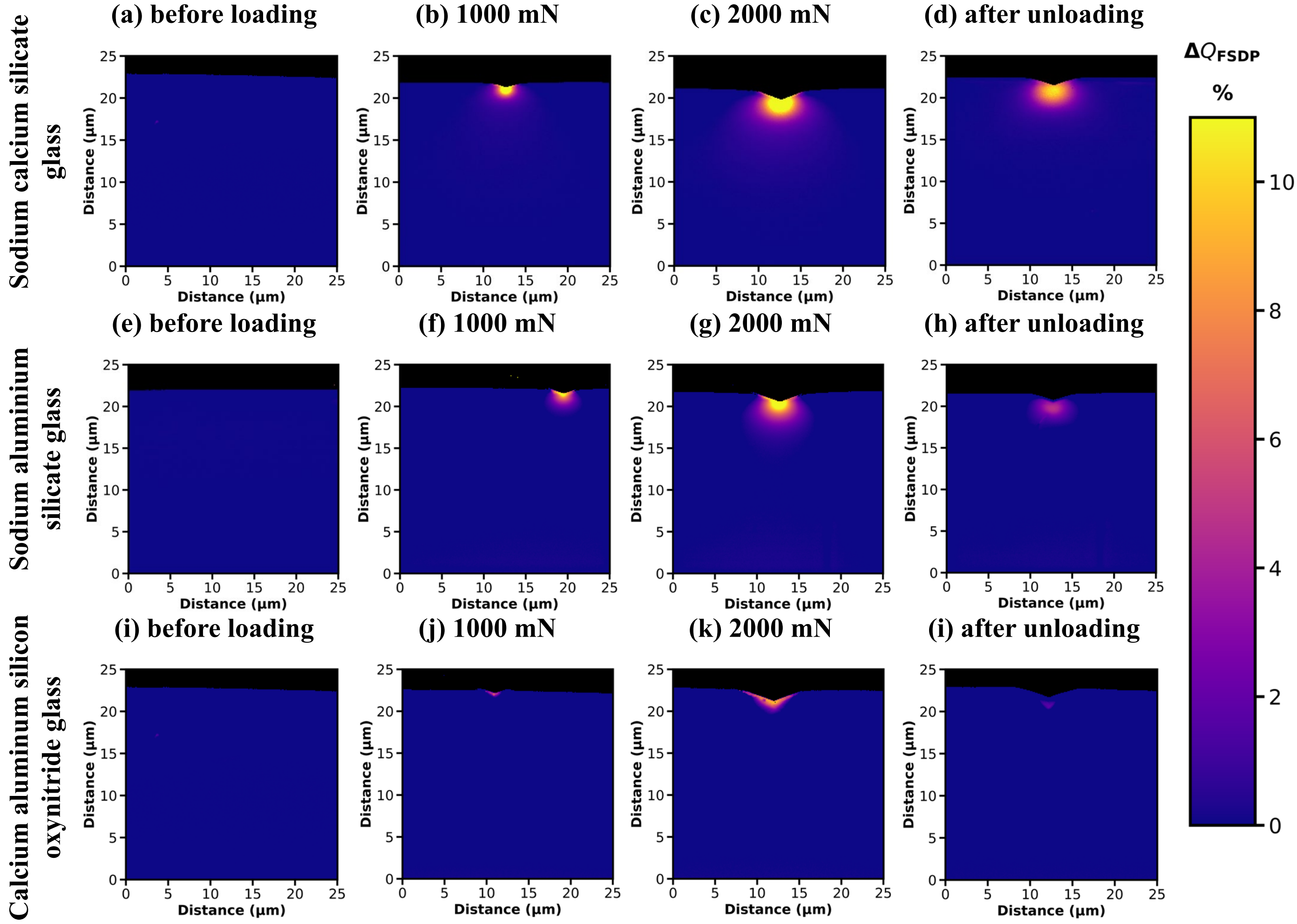

**Figure 5.** Deformation maps obtained over different indentation loading cycles for (a-d) NCS, (e-h) NAS, and (i-l) oxynitride glasses at different stages of the indentation process (before loading, 1 N load, 2 N load, and after unloading). All the 2D maps have the same normalized color scales with respect to the initial undeformed FSDP position value, allowing for direct comparison across samples.

To analyze the deformation zones in more details, Figure 6 shows the stacked $I(q)$ curves obtained along the dashed line drawn on the corresponding 2D maps for the four different glasses used in this study. As the glasses are indented, the FSDP position shifts to a higher $q$ value for all glasses, while the intensity of the FSDP decreases. For NCS and NAS glasses, two overlapping peaks can be seen in the FSDP region, while their relative contribution changing upon indentation. More specifically, during indentation, the low-$q$ peak (~1.6 Å$^{-1}$) transforms into a higher-$q$ peak (~2.2 Å$^{-1}$). The CAS-N glass features a unimodal distribution of FSDP, similar to that of vitreous silica. The maximum shift in $q_{FSDP}$ achieved at 2 N load is 1.743 Å$^{-1}$ ~13% higher than the $q_{FSDP}$ of the undeformed portion of 1.540 Å$^{-1}$) for vitreous silica, whereas it is 1.900 Å$^{-1}$ for NCS (17% higher than its initial value of 1.625 Å$^{-1}$), 1.956 Å$^{-1}$ for NAS (16% higher than its initial value of 1.692 Å$^{-1}$), and 2.223 Å$^{-1}$ for CAS-N (10% higher than its initial value of 2.018 Å$^{-1}$). To further understand this, we note that Poisson's ratio is one of the macroscopic indicators of the structural packing in the glass network . As network modifiers (alkaline and alkaline earths) are introduced into silica glass, $q_{\text{FSDP}}$ increases due to decreasing free volume, i.e., $q_{\text{FSDP}}$ is positively correlated with Poisson's ratio in the present glasses. The smaller relative FSDP shift observed for CAS-N, together with its higher initial $q_{\text{FSDP}}$ and higher Poisson's ratio, suggests a more densely packed and less densification-prone network compared with the oxide glasses.

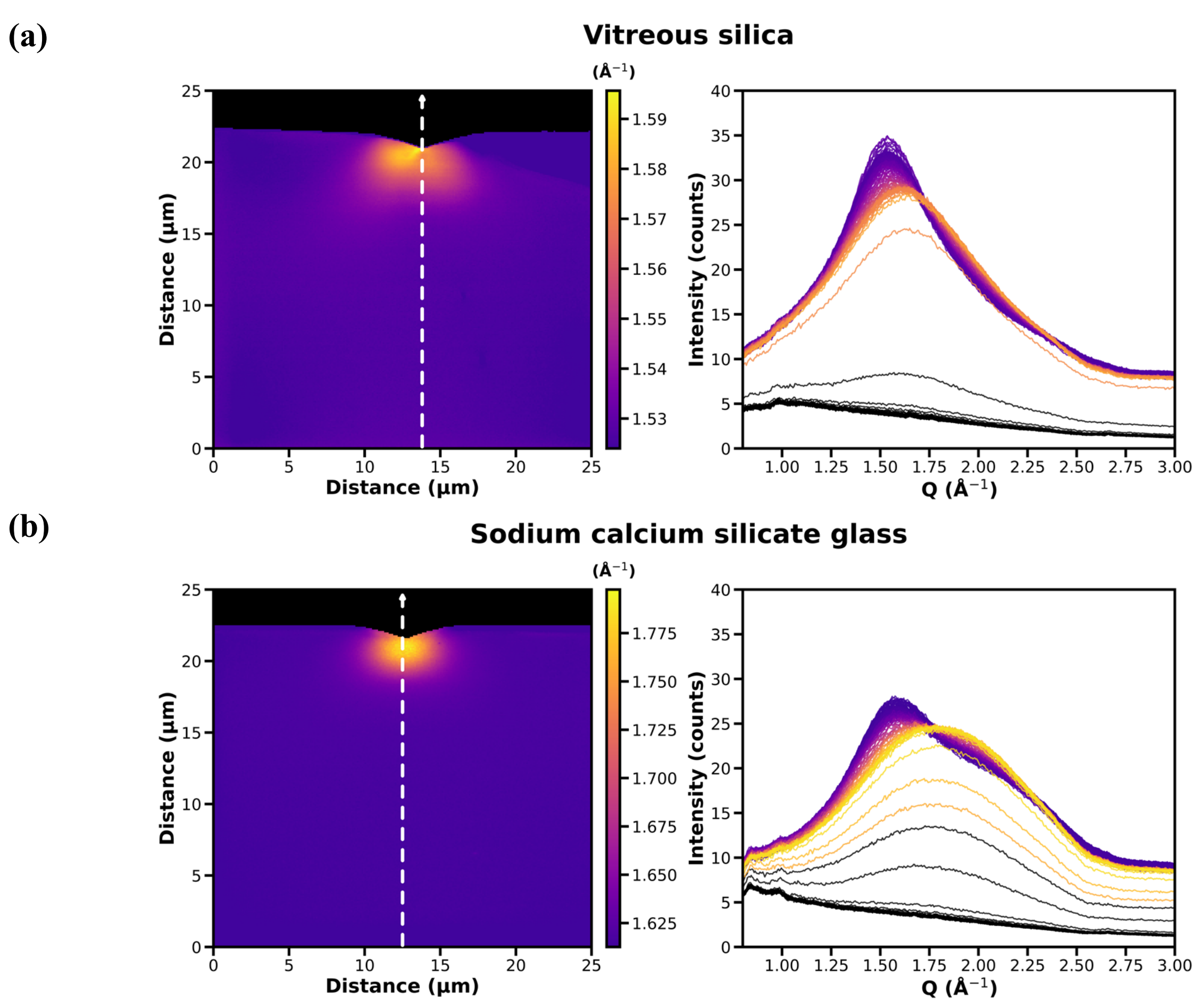

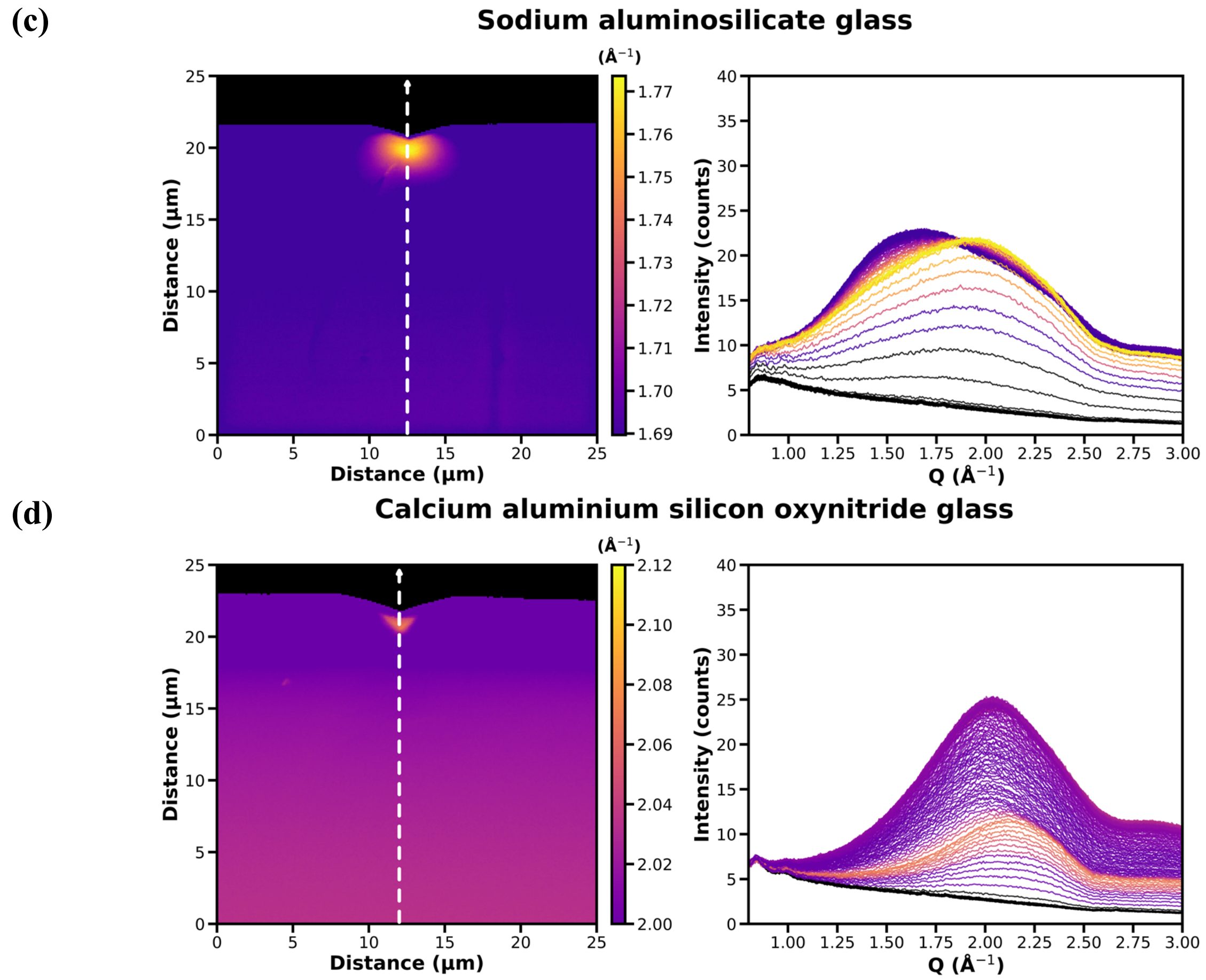

**Figure 6.** Stacked $I(q)$ curves obtained along the dashed lines drawn on the 2D maps for (a) vitreous silica, (b) soda lime silicate glass, (c) sodium aluminosilicate glass, and (d) oxynitride glass at maximum load of 2 N.

Intensity maps for all glasses are shown in Supporting Figure S12, from which various features that scatter the incoming X-rays can be clearly seen. These features include the edges of the indenter tip, sample surface, and cracks formed during indentation. Notably, vitreous silica does not feature any cracking at 1 N, but a cone-type crack appears near one side of the indenter face at 2 N load. The NCS glass features median cracking at a load as low as 1 N (Supporting Figure S12f) and the crack grows with increasing load (Supporting Figure S12g) as well as after unloading (Supporting Figure S12h). The NAS and CAS-N glasses only cracked (median crack) after complete unloading. We also note that the cracking nature of the oxynitride glass is different from that of the oxide glasses, i.e., the crack passes through the apparent shear band and gets deviated (Supporting Figure S12p).

Figure 7 and Supporting Figure S13 show the contour plots after segmentation for the different glasses at various loading conditions. These plots isolate the deformation zone into distinct islands, clearly showing how far the deformation penetrates the glass. This approach allows for a direct comparison of deformation behavior between different glasses under the same load, or for the same glass at varying loads. As a general trend, as the indentation load increases from 1 to 2 N, the size of the deformation zone expands. The deformation zone decreases in size upon complete unloading in comparison to the zone at 2 N for all glasses. This is as expected based on the elastic recovery upon unloading. The size (cross-section area) of the deformation zone at 1 N load is found to be 59 $\mu m^2$ for vitreous silica, 58 $\mu m^2$ for NCS, 17 $\mu m^2$ for NAS glasses, and 0.8 $\mu m^2$ for CAS-N glasses. Upon increasing the load to 2 N, the zone size increases but disproportionately among the different glasses, i.e., it is 99, 114, 53, and 8 $\mu m^2$ for silica, NCS, NAS, and CAS-N glasses, respectively. As this zone is comprised of several deformation aspects, the highly densifying silica glass features a deformation zone with varying shape upon the application of load along with the elastic strain field. However, an increasing load tends to densify the glass locally, which explains the disproportionate change in the size of the deformation zone with increasing load among the different glasses. For the NCS glass, densification occurs concurrently with shear during the initial stage of loading. Particularly, the deformation zone size, which is smaller than for vitreous silica at 1 N, becomes larger than vitreous silica at 2 N. The predominant mode of deformation changes from mixed mode to highly densifying with increasing load, in agreement with previous reports on the load dependence of densification in soda-lime silicate glass [8]. As there is negligible NBO concentration per tetrahedron (0.036) in the present NAS glass, this glass network is highly connected and hence no shear flow could be discernible from the indentation. The size of the plastic deformation zone after unloading from 2 N is found to be 58 $\mu m^2$ for silica, 77 $\mu m^2$ for NCS, 20 $\mu m^2$ for NAS, and 1.1 $\mu m^2$ for CAS-N.

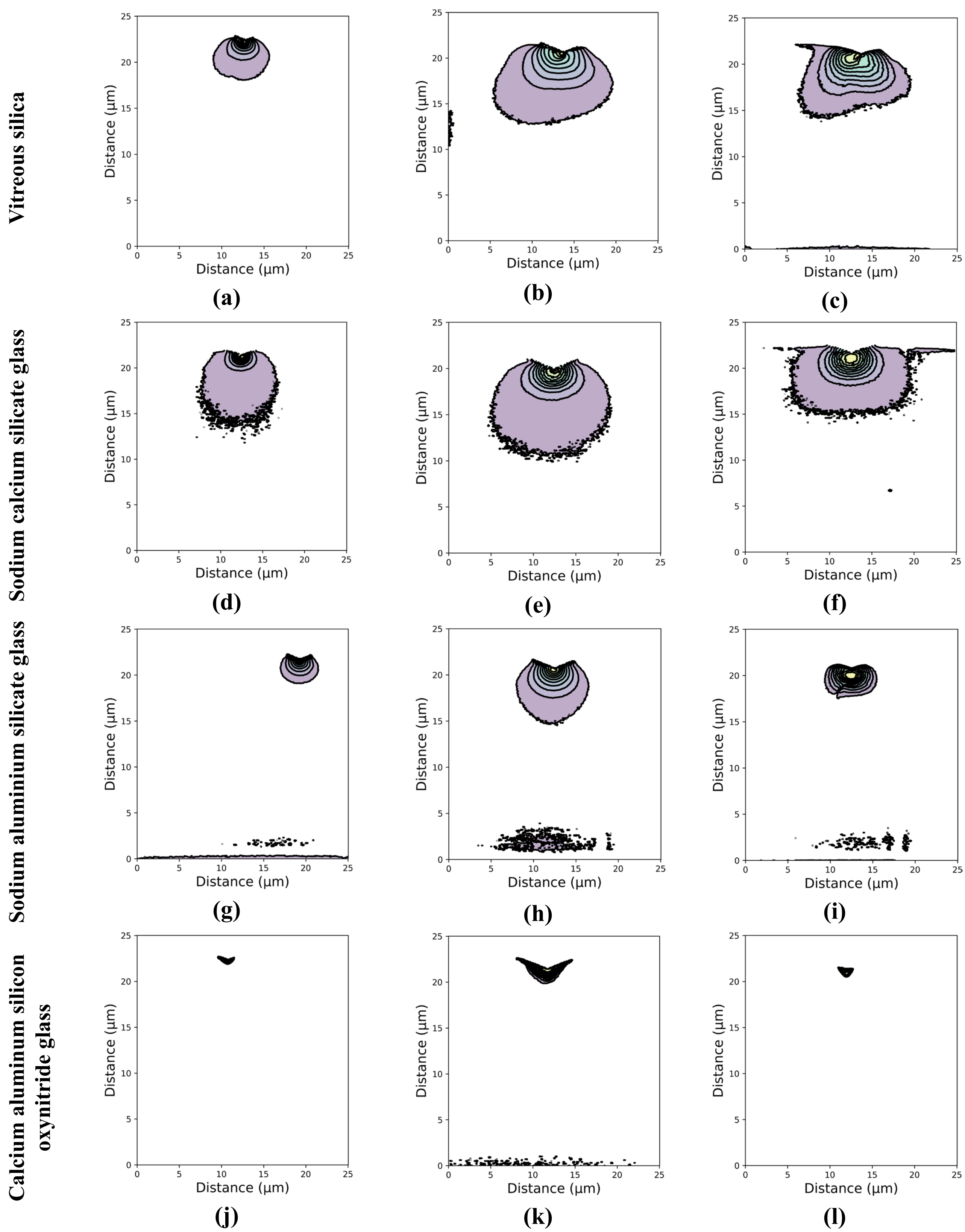

**Figure 7.** Isolated contour plots, showing regions of continuous deformation (sub-contours) based on FSDP position values after thresholding. Left column panels show glasses at the maximum load (2 N), while right

column panels show samples after unloading from this load. Results are shown for (a-c) vitreous silica, (d-f) NCS, (g-i) NAS, and (j-l) CAS-N glasses.

The relative change in deformation from the maximum applied load to complete removal of load reflects the partial recovery of the deformation, implying the elastic deformation (strain field). This also allows us to derive the remanent plastic deformation, which provides the densification map for a predominantly densifying glass such as vitreous silica. Based on this analysis, the densification for silica accounts for ~59% of the total deformation incurred at the maximum load of 2 N using the 136° diamond wedge tip. The recovered 41% of the deformation upon unloading is ascribed to elastic deformation as well as some extent of shear deformation, although shear flow should be small in the case of silica. The unrecovered deformation area accounts for 67% in NCS, 37% in NAS, and 13% in CAS-N glasses. The remaining parts of the deformation must then correspond to a combination of elastic strain, shear flow, and shear banding.

## 4. Conclusions

We have used *in situ* synchrotron X-ray nanoscattering with a spatial resolution of 100 nm and a 136° wedge indenter tip to map the evolution of indentation-induced deformation in four oxide and oxynitride glasses during loading and after unloading. With this technique we are able to resolve and distinguish elastic from plastic contributions in addition to visualizing densification, shear flow, shear banding, and cracking. The results show that both the size and morphology of the sub-surface deformation zone depend strongly on glass composition and, in particular, on Poisson's ratio. For a given glass, the deformation zone evolves continuously throughout the indentation cycle, reflecting changes in the balance between elastic recovery, permanent densification, and shear-related deformation. Among the studied compositions, the CAS-N oxynitride glass showed the most localized deformation zone and the smallest residual deformation after unloading, consistent with its high Poisson's ratio and densely packed network. By combining FSDP-based structural mapping with threshold-based segmentation, we thus demonstrate a non-destructive route to quantify the deformation zone directly, without post-indentation annealing, thereby preserving the intrinsic deformation state of the glass. This approach also provides new insight into how local medium-range structural rearrangements govern indentation response and residual deformation in different glass families. Our findings therefore offer a framework for linking composition, structure, and contact-damage behavior, supporting the future design of glasses with improved resistance to flaw formation and mechanical failure. In addition, the present methodology provides an experimental basis for validating deformation models of glasses under sharp contact.

**Acknowledgements**

We acknowledge Randall E. Youngman (Corning Inc.) for providing silica samples and Henning F. Poulsen (Technical University of Denmark) for helpful discussion. This work was supported through from the ESS lighthouse on hard materials in 3D, SOLID, funded by the Danish Agency for Science and Higher Education, grant number 8144-00002B as well as the European Union (ERC, NewGLASS, 101044664). We acknowledge the MAX IV Laboratory for beamtime on the NanoMAX beamline under proposal 20231252 & 20230857. Research conducted at MAX IV, a Swedish national user facility, is supported by Vetenskapsrådet (Swedish Research Council, VR) under contract 2018-07152, Vinnova (Swedish Governmental Agency for Innovation Systems) under contract 2018-04969 and Formas under contract 2019-02496. We acknowledge the travel support provided through DanScatt by the Danish Agency for Science, Technology, and Innovation for this beamtime experiment.

**References**

[1] L. Wondraczek, E. Bouchbinder, A. Ehrlicher, J.C. Mauro, R. Sajzew, M.M. Smedskjaer, Advancing the mechanical performance of glasses: perspectives and challenges, Advanced Materials 34(14) (2022) 2109029.
[2] I.W. Donald, Methods for improving the mechanical properties of oxide glasses, Journal of Materials Science 24(12) (1989) 4177-4208.
[3] T. Rouxel, S. Yoshida, The fracture toughness of inorganic glasses, Journal of the American Ceramic Society 100(10) (2017) 4374-4396.
[4] A.S. Argon, Inelastic deformation and fracture in oxide, metallic, and polymeric glasses, in: D.R. Uhlmann, N.J. Kreidl (Eds.), Glass Science and Technology, Elsevier1980, pp. 79-132.
[5] S. Kasimuthumaniyan, S. Sahoo, M.M. Smedskjaer, N.M.A. Krishnan, N.N. Gosvami, Quantifying the densification and shear flow under indentation deformation in borosilicate glasses, International Journal of Applied Glass Science 13(4) (2022) 526-538.
[6] K. Januchta, M.M. Smedskjaer, Indentation deformation in oxide glasses: Quantification, structural changes, and relation to cracking, Journal of Non-Crystalline Solids: X 1 (2019) 100007.
[7] T. Rouxel, J.-i. Jang, U. Ramamurty, Indentation of glasses, Progress in Materials Science 121 (2021) 100834.
[8] S. Yoshida, J.-C. Sanglebœuf, T. Rouxel, Quantitative evaluation of indentation-induced densification in glass, Journal of Materials Research 20(12) (2005) 3404-3412.
[9] J.-P. Guin, K. Han, L. Charleux, J.-C. Sangleboeuf, M. Ferry, V. Keryvin, A new nanometre resolution method for probing densification ratio at nanoindentation sites in glass: Unravelling discrepancies in the literature, Acta Materialia 274 (2024) 120005.
[10] A. Baggott, M. Mazaheri, B.J. Inkson, 3D characterisation of indentation induced sub-surface cracking in silicon nitride using FIB tomography, Journal of the European Ceramic Society 39(13) (2019) 3620-3626.

[11] N. Cuadrado, J. Seuba, D. Casellas, M. Anglada, E. Jiménez-Piqué, Geometry of nanoindentation cube-corner cracks observed by FIB tomography: Implication for fracture resistance estimation, Journal of the European Ceramic Society 35(10) (2015) 2949-2955.
[12] A. Perriot, D. Vandembroucq, E. Barthel, V. Martinez, L. Grosvalet, C. Martinet, B. Champagnon, Raman microspectroscopic characterization of amorphous silica plastic behavior, Journal of the American Ceramic Society 89(2) (2006) 596-601.
[13] T. Deschamps, C. Martinet, J. Bruneel, B. Champagnon, Soda-lime silicate glass under hydrostatic pressure and indentation: a micro-Raman study, Journal of Physics: Condensed Matter 23(3) (2011) 035402.
[14] S. Bruns, T. Uesbeck, S. Fuhrmann, M. Tarragó Aymerich, L. Wondraczek, D. de Ligny, K. Durst, Indentation densification of fused silica assessed by raman spectroscopy and constitutive finite element analysis, Journal of the American Ceramic Society 103(5) (2020) 3076-3088.
[15] H. Tran, S. Clement, R. Vialla, D. Vandembroucq, B. Ruffle, Micro-Brillouin spectroscopy mapping of the residual density field induced by Vickers indentation in a soda-lime silicate glass, Applied Physics Letters 100(23) (2012).
[16] A. Berthelot, E. Romeo, X. Dagany, T. Deschamps, E. Herry, R. Kineider, Y. De Leon, E. Brassac, G. Kermouche, E. Barthel, Strength of Brillouin spectroscopy to identify spatial densification model in indented silica, Journal of Non-Crystalline Solids 639 (2024) 123058.
[17] H. Liu, B. Deng, S. Sundararaman, Y. Shi, L. Huang, Understanding the response of aluminosilicate and aluminoborate glasses to sharp contact loading using molecular dynamics simulation, Journal of Applied Physics 128(3) (2020).
[18] J.F. Christensen, M.F.U. Jalaludeen, S.S. Sørensen, A.K. Christensen, X. Ge, T. Du, Y. Yue, A. Davydok, C. Krywka, L. Wondraczek, H.F. Poulsen, M.M. Smedskjaer, In situ mapping of indentation-induced densification and cracking in vitreous silica by nanofocus X-ray scattering, arXiv preprint arXiv:.05699 (2025).
[19] S. Fuhrmann, G.N.B.M. de Macedo, R. Limbach, C. Krywka, S. Bruns, K. Durst, L. Wondraczek, Indentation-induced structural changes in vitreous silica probed by in-situ small-angle X-ray scattering, Frontiers in Materials Volume 7 - 2020 (2020).
[20] S. Yoshida, T.H. Nguyen, A. Yamada, J. Matsuoka, In-situ Raman measurements of silicate glasses during Vickers indentation, Materials transactions 60(8) (2019) 1428-1432.
[21] Y.B. Gerbig, C.A. Michaels, In-situ Raman spectroscopic measurements of the deformation region in indented glasses, Journal of Non-Crystalline Solids 530 (2020) 119828.
[22] S. Yoshida, Indentation deformation and cracking in oxide glass–toward understanding of crack nucleation, Journal of Non-Crystalline Solids: X 1 (2019) 100009.
[23] G.A. Rosales-Sosa, E. Barthel, Y. Kato, M. Bourguignon, A. Yamada, T. Inoue, S. Nakane, H. Yamazaki, Indentation stress fields in brittle materials: A micro-photoelastic investigation in silicate glasses, Acta Materialia (2025) 120973.
[24] G. Okuma, K. Maeda, S. Yoshida, A. Takeuchi, F. Wakai, Morphology of subsurface cracks in glass-ceramics induced by Vickers indentation observed by synchrotron X-ray multiscale tomography, Scientific reports 12(1) (2022) 6994.
[25] H. He, Z. Chen, Y.-T. Lin, S.H. Hahn, J. Yu, A.C.T. van Duin, T.D. Gokus, S.V. Rotkin, S.H. Kim, Subsurface structural change of silica upon nanoscale physical contact: Chemical plasticity beyond topographic elasticity, Acta Materialia 208 (2021) 116694.
[26] J. Gamcová, G. Mohanty, Š. Michalik, J. Wehrs, J. Bednarčík, C. Krywka, J.M. Breguet, J. Michler, H. Franz, Mapping strain fields induced in Zr-based bulk metallic glasses during in-situ nanoindentation by X-ray nanodiffraction, Applied Physics Letters 108(3) (2016).
[27] H.F. Poulsen, J.A. Wert, J. Neuefeind, V. Honkimäki, M. Daymond, Measuring strain distributions in amorphous materials, Nature Materials 4(1) (2005) 33-36.

[28] R. Shi, H. Tanaka, Distinct signature of local tetrahedral ordering in the scattering function of covalent liquids and glasses, Science Advances 5(3) (2019) eaav3194.
[29] S. Moss, D. Price, Random packing of structural units and the first sharp diffraction peak in glasses, Physics of disordered materials, Springer1985, pp. 77-95.
[30] P. Gaskell, D. Wallis, Medium-range order in silica, the canonical network glass, Physical review letters 76(1) (1996) 66.
[31] M.T. Shatnawi, The first sharp diffraction peak in the total structure function of amorphous chalcogenide glasses: anomalous characteristics and controversial views, New Journal of Glass Ceramics 6(03) (2016) 37.
[32] S. Vepřek, H. Beyeler, On the interpretation of the first, sharp maximum in the X-ray scattering pattern of non-crystalline solids and liquids, Philosophical Magazine B 44(5) (1981) 557-567.
[33] Y. Shi, N.T. Lonnroth, R.E. Youngman, S.J. Rzoska, M. Bockowski, M.M. Smedskjaer, Pressure-induced structural changes in titanophosphate glasses studied by neutron and X-ray total scattering analyses, Journal of Non-Crystalline Solids 483 (2018) 50-59.
[34] D. Adler, Physics of disordered materials, Springer Science & Business Media 2012.
[35] Y. Shi, J. Neuefeind, D. Ma, K. Page, L.A. Lamberson, N.J. Smith, A. Tandia, A.P. Song, Ring size distribution in silicate glasses revealed by neutron scattering first sharp diffraction peak analysis, Journal of Non-Crystalline Solids 516 (2019) 71-81.
[36] U. Johansson, D. Carbone, S. Kalbfleisch, A. Bjorling, M. Kahnt, S. Sala, T. Stankevic, M. Liebi, A. Rodriguez Fernandez, B. Bring, D. Paterson, K. Thanell, P. Bell, D. Erb, C. Weninger, Z. Matej, L. Roslund, K. Ahnberg, B. Norsk Jensen, H. Tarawneh, A. Mikkelsen, U. Vogt, NanoMAX: the hard X-ray nanoprobe beamline at the MAX IV Laboratory, Journal of Synchrotron Radiation 28(6) (2021) 1935-1947.
[37] T. Rouxel, H. Ji, J. Guin, F. Augereau, B. Rufflé, Indentation deformation mechanism in glass: densification versus shear flow, Journal of Applied Physics 107(9) (2010).
[38] T. Rouxel, H. Ji, T. Hammouda, A. Moréac, Poisson's ratio and the densification of glass under high pressure, Physical Review Letters 100(22) (2008) 225501.
[39] H. Corning, 7979, 7980, 8655 Fused Silica–Optical Materials Product Information, Corning–Specialty Materials, 2015.
[40] S. Ali, A. Ellison, J. Luo, M. Edén, Composition–structure–property relationships of transparent Ca–Al–Si–O–N oxynitride glasses: the roles of nitrogen and aluminum, Journal of the American Ceramic Society 106(3) (2023) 1748-1765.
[41] M.F.U. Jalaludeen, X-ray Attenuation Calculator for Oxide glasses, Software Release, Zenodo, 2025.
[42] G. Lotze, A.H.S. Iyer, O. Backe, S. Kalbfleisch, M.H. Colliander, In situ characterization of stresses, deformation and fracture of thin films using transmission X-ray nanodiffraction microscopy, Journal of Synchrotron Radiation 31(1) (2024) 42-54.
[43] A. Björling, S. Kalbfleisch, M. Kahnt, S. Sala, K. Parfeniukas, U. Vogt, D. Carbone, U. Johansson, Ptychographic characterization of a coherent nanofocused X-ray beam, Opt. Express 28(4) (2020) 5069-5076.
[44] D. Carbone, S. Kalbfleisch, U. Johansson, A. Bjorling, M. Kahnt, S. Sala, T. Stankevic, A. Rodriguez-Fernandez, B. Bring, Z. Matej, P. Bell, D. Erb, V. Hardion, C. Weninger, H. Al-Sallami, J. Lidon-Simon, S. Carlson, A. Jerrebo, B. Norsk Jensen, A. Bjermo, K. Ahnberg, L. Roslund, Design and performance of a dedicated coherent X-ray scanning diffraction instrument at beamline NanoMAX of MAX IV, Journal of Synchrotron Radiation 29(3) (2022) 876-887.
[45] L. Lurio, N. Mulders, M. Paetkau, P.R. Jemian, S. Narayanan, A. Sandy, Windows for small-angle X-ray scattering cryostats, Journal of Synchrotron Radiation 14(6) (2007) 527-531.

[46] W. Bras, H. Stanley, Unexpected effects in non crystalline materials exposed to X-ray radiation, Journal of Non-Crystalline Solids 451 (2016) 153-160.
[47] G. Ashiotis, A. Deschildre, Z. Nawaz, J.P. Wright, D. Karkoulis, F.E. Picca, J. Kieffer, The fast azimuthal integration Python library: pyFAI, Applied Crystallography 48(2) (2015) 510-519.
[48] A.B. Jensen, T.E.K. Christensen, C. Weninger, H. Birkedal, Very large-scale diffraction investigations enabled by a matrix-multiplication facilitated radial and azimuthal integration algorithm: MatFRAIA, Journal of Synchrotron Radiation 29(6) (2022) 1420-1428.
[49] W.C. Oliver, G.M. Pharr, Measurement of hardness and elastic modulus by instrumented indentation: Advances in understanding and refinements to methodology, Journal of Materials Research 19(1) (2004) 3-20.
[50] G.N. Greaves, A.L. Greer, R.S. Lakes, T. Rouxel, Poisson's ratio and modern materials, Nature Materials 10(11) (2011) 823-837.

## Supporting Information

***for***

# Uncovering the deformation mechanism of glasses during indentation through high-resolution X-ray scattering

*M. Faizal Ussama Jalaludeen*[1]*, Søren S. Sørensen*[1]*, Johan F. S. Christensen*[1]*, Anders K. R. Christensen*[1]*, Sidsel M. Johansen*[1]*, Samraj Mollick*[1]*, Yuanzheng Yue*[1]*, Sharafat Ali*[2]*, Sebastian Kalbfleisch*[3]*, Morten M. Smedskjaer*[1,*]

[1]Department of Chemistry and Bioscience, Aalborg University, DK-9220 Aalborg, Denmark

[2]Department of Built Environment and Energy Technology, Linnæus University, SE-351 95 Växjö, Sweden

[3]MAX IV Laboratory, Lund University, PO Box 118, S-221 00 Lund, Sweden

*Corresponding Author. Email: mos@bio.aau.dk

## Supporting Text S1: Beam-induced damage

An important aspect to consider is the structural stability of the glasses to the X-ray beam when exposed for longer time or scanned multiple times on the same area. We find that the beam-induced damage is highly material dependent, i.e., some glasses are more prone to beam-induced structural damage than others. As shown in Supporting Figure S9 for vitreous silica, the homogenous area scanned during indentation process far away from the indentation deformation zone has a slightly different FSDP (hence different atomic-scale structure) than the larger overview area. This shows that this glass structure changes upon beam exposure. That is, as we expose these glasses for longer time, the structure will no more be the same as that of the starting structure. In our case, as we scan the sample over the same area for increasing indentation loads, the structure might have changed just by the X-ray beam, in addition to the indentation.

To understand the structural differences between areas exposed for different times, we also performed repeated area scans in a following pattern (Supporting Figure S10). We find that in vitreous silica there is a systematic shift in the FSDP position from their initial $q$ position with increasing exposure time of the X-rays on the sample. The longer the exposure time, the more observed damage in the glass. This phenomenon is, however, composition dependent as only vitreous silica is prone to structural changes upon irradiation, as no noticeable changes are observed in other glasses within the probed exposure time up to 250 ms (five repetitions of 50 ms exposure time each) at 14 keV.

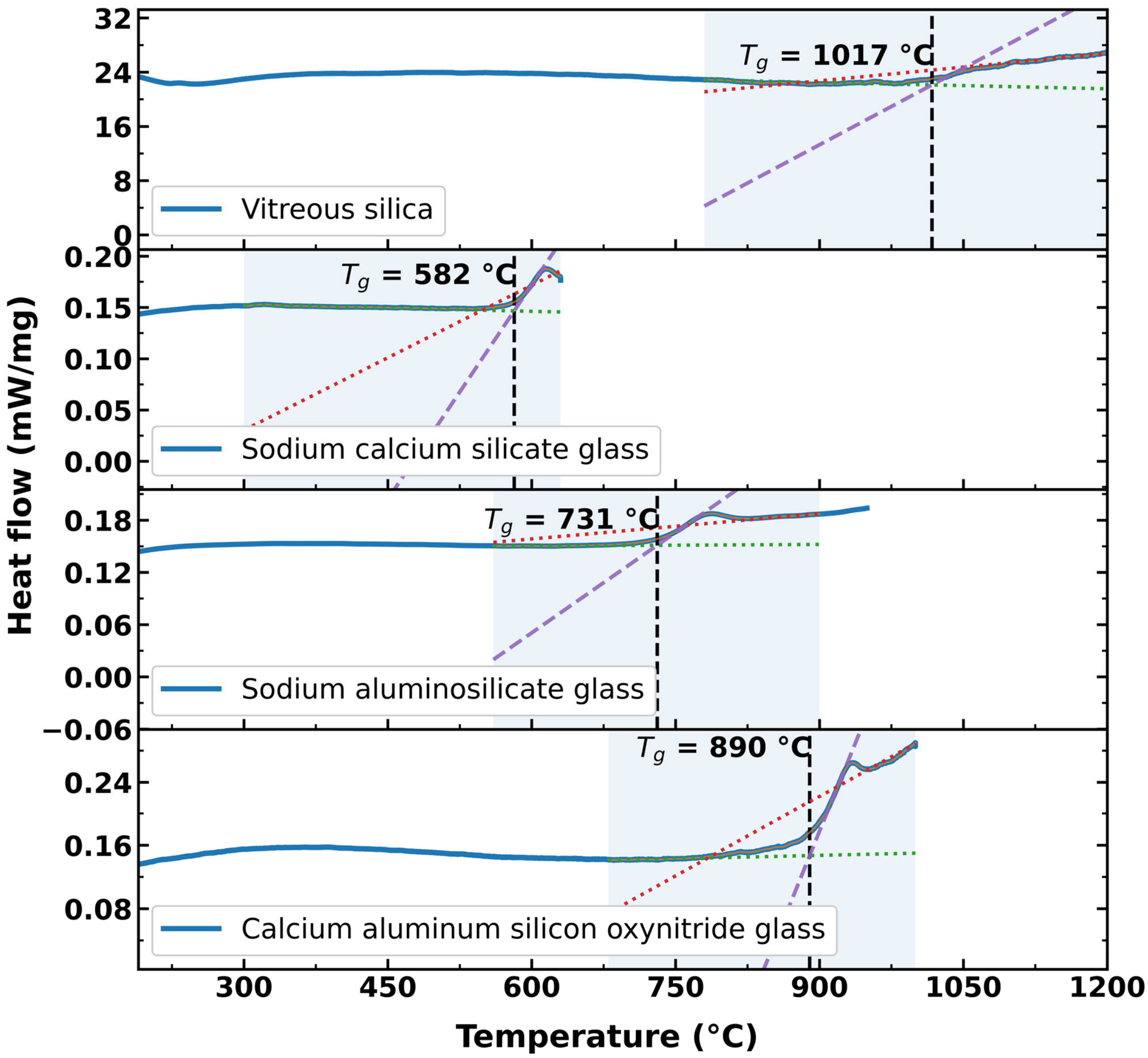

**Figure S1.** DSC curves of different oxide and oxynitride glasses, with endotherm pointing upwards. The heating rate used for DSC measurement was 20 °C/min for vitreous silica, whereas it was 10 °C/min for the other glasses.

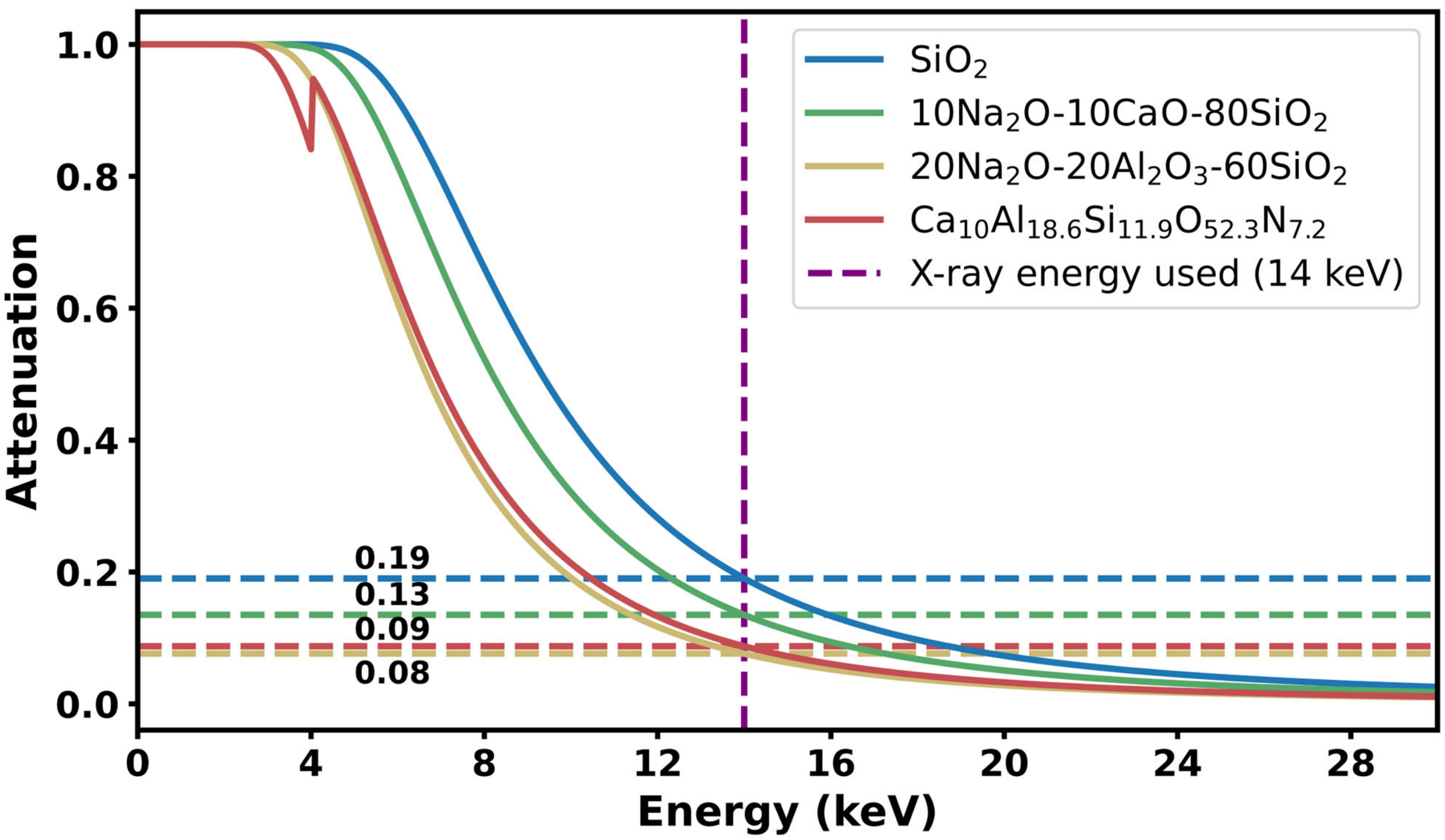

**Figure S2.** X-ray interaction cross section curves over wide energy ranges for four different glasses of ~135 µm thickness used in this study. Vertical purple line highlights the energy used in this study (14 keV) and the horizontal lines indicate the attenuation values for each of the oxide glasses at this X-ray energy.

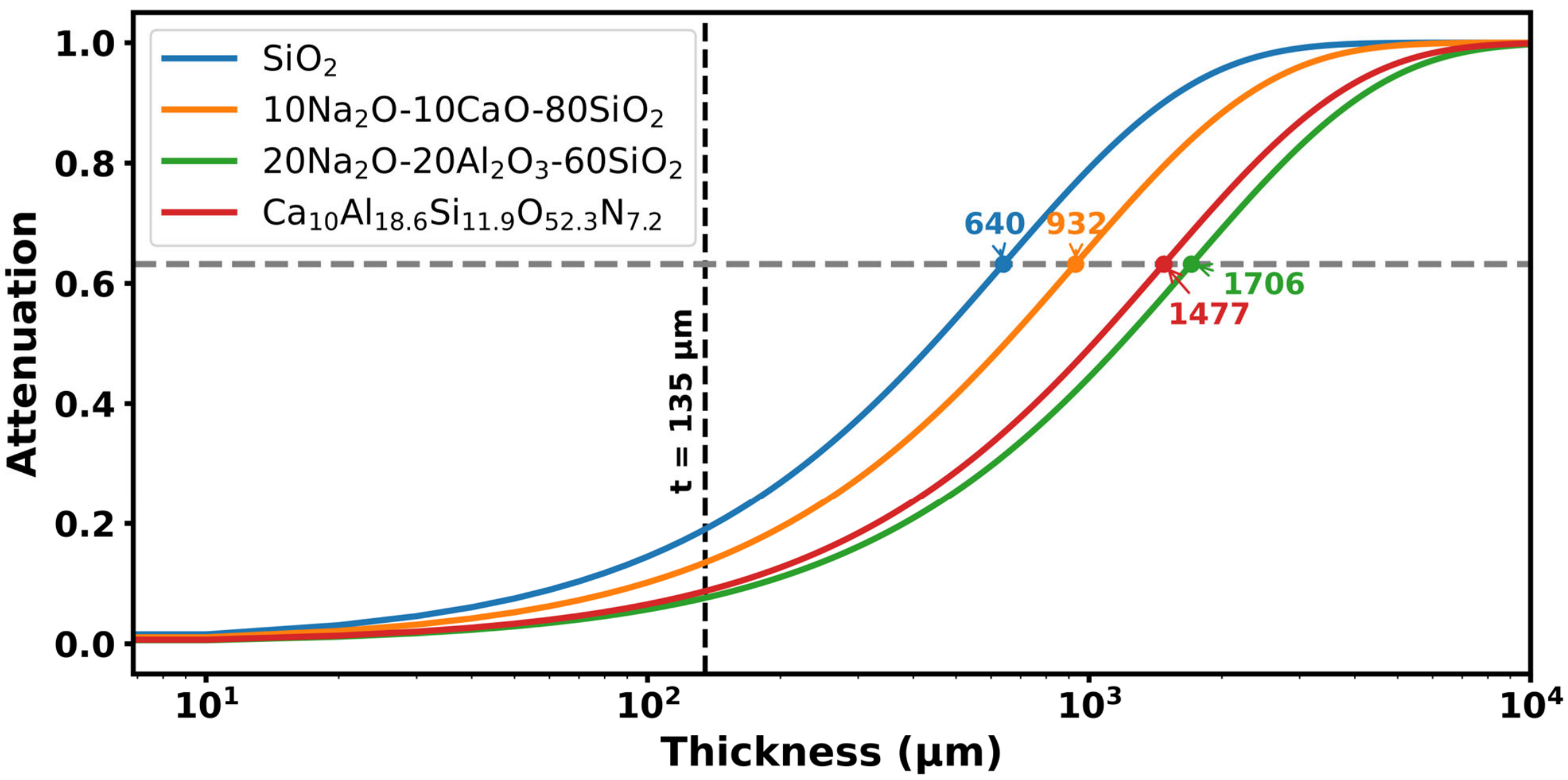

**Figure S3.** Attenuation of 14 keV X-rays by the four different glasses as a function of thickness. Vertical black line refers the thickness chosen in this study, i.e., ~135 μm. The horizontal line at 0.632 of the total attenutation corresponds to the 1/e depth, and the values for each of the oxide glasses at a specific X-ray energy (14 keV) is annotated next to the respective curves.

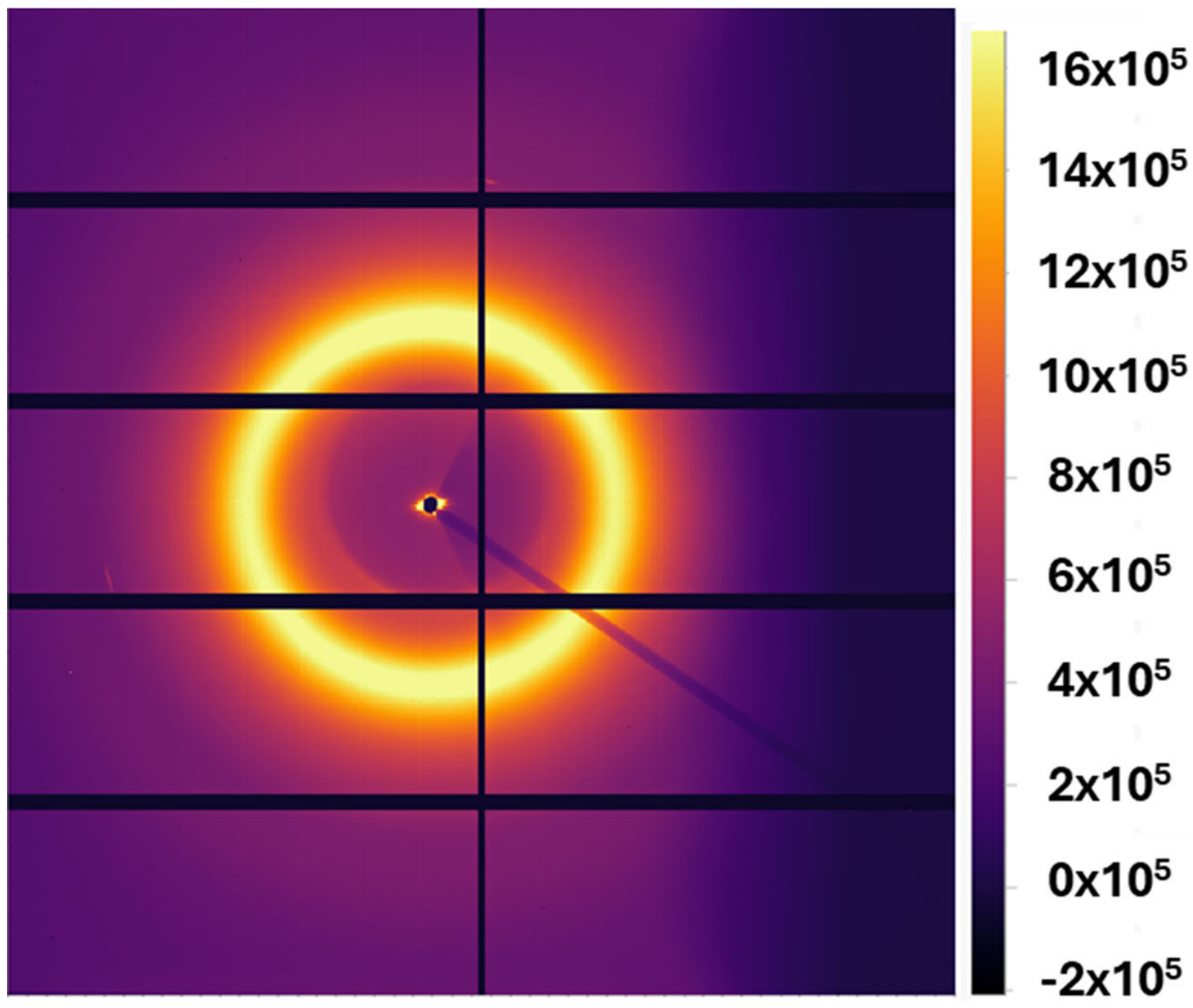

**Figure S4.** Example of raw X-ray intensities measured on the Pilatus3 1M detector placed at a distance 150 mm away from the sample. The shadow effect, which is caused by the indenter tip, can be seen near the beamstop.

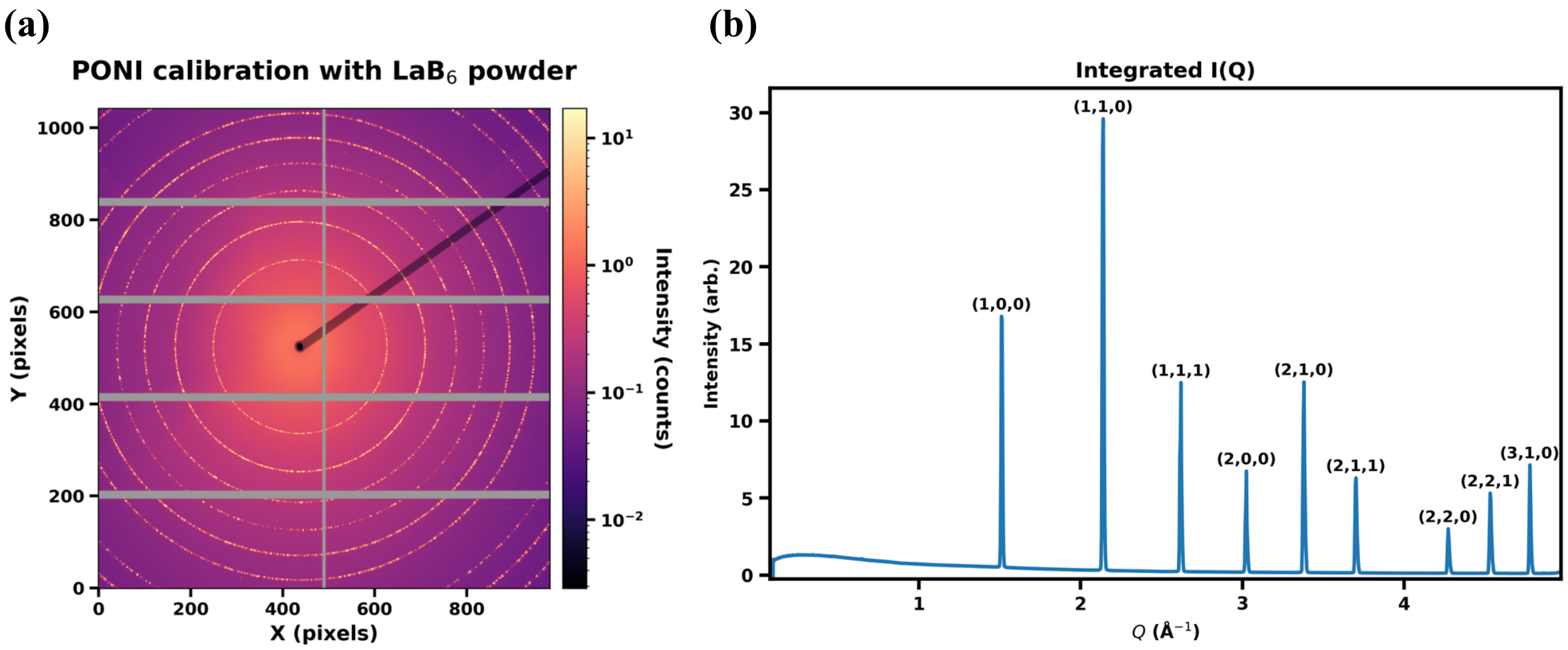

**Figure S5.** (a) Raw detector image as acquired from the calibrant, lanthanum hexaboride ($LaB_6$), with mask applied (black [beam stop and dead pixels] and grey regions [inter-module/detector gaps]). This was used to obtain the point of normal incidence (PONI) at the operating beam conditions (14 keV), as well the sample to detector distance (SDD) for further processing. (b) Integrated 1D intensity curve $I(q)$ with lattice planes indexed after calibration obtained using the $LaB_6$ calibrant.

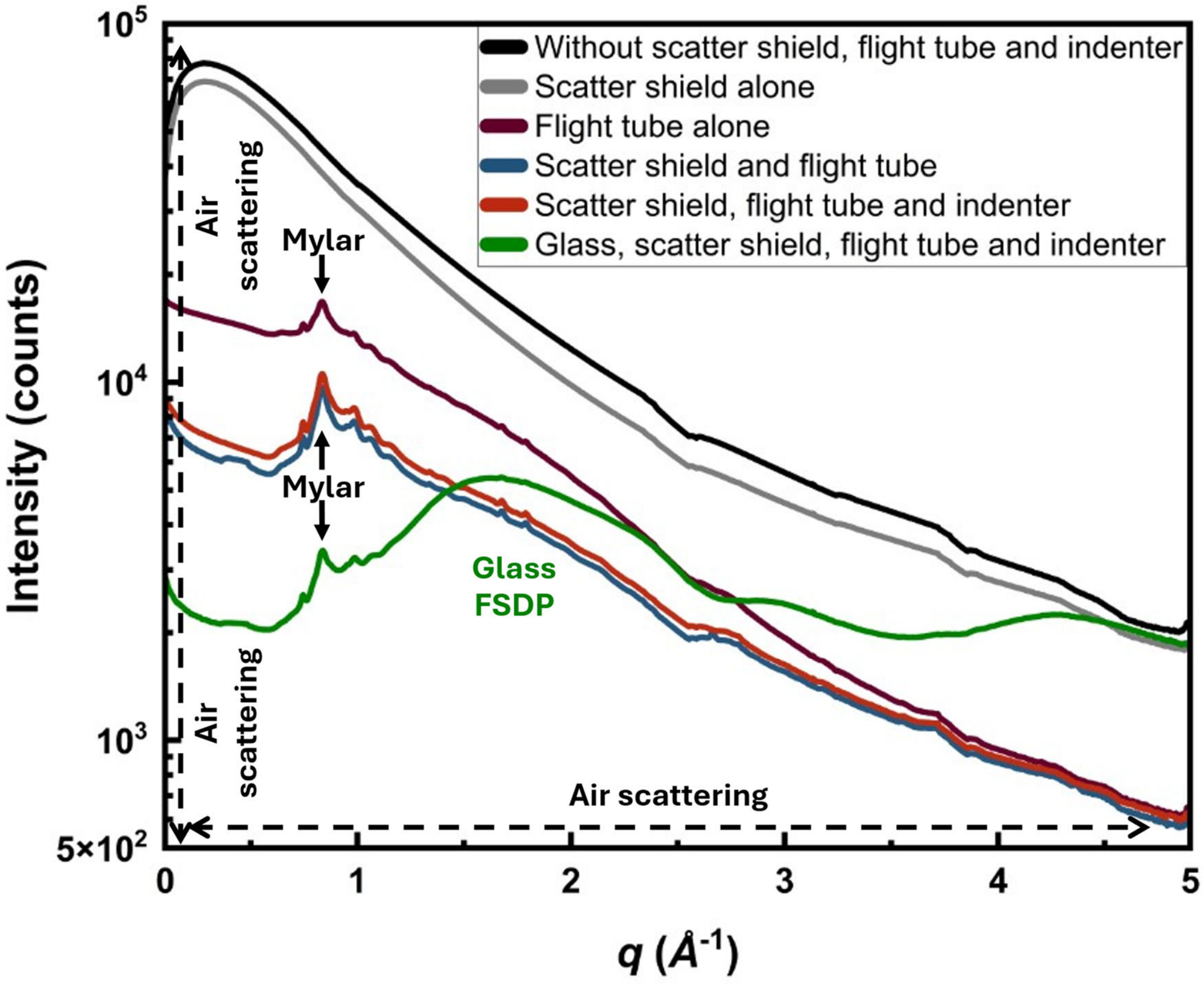

**Figure S6.** Raw scattering signal detected in presence and absence of various combinations of the components and their suppression of air scattering. An example scattering signal from NAS-60 glass in presence of all components is provided to highlight the effectiveness of these additional components in suppressing the air scattering.

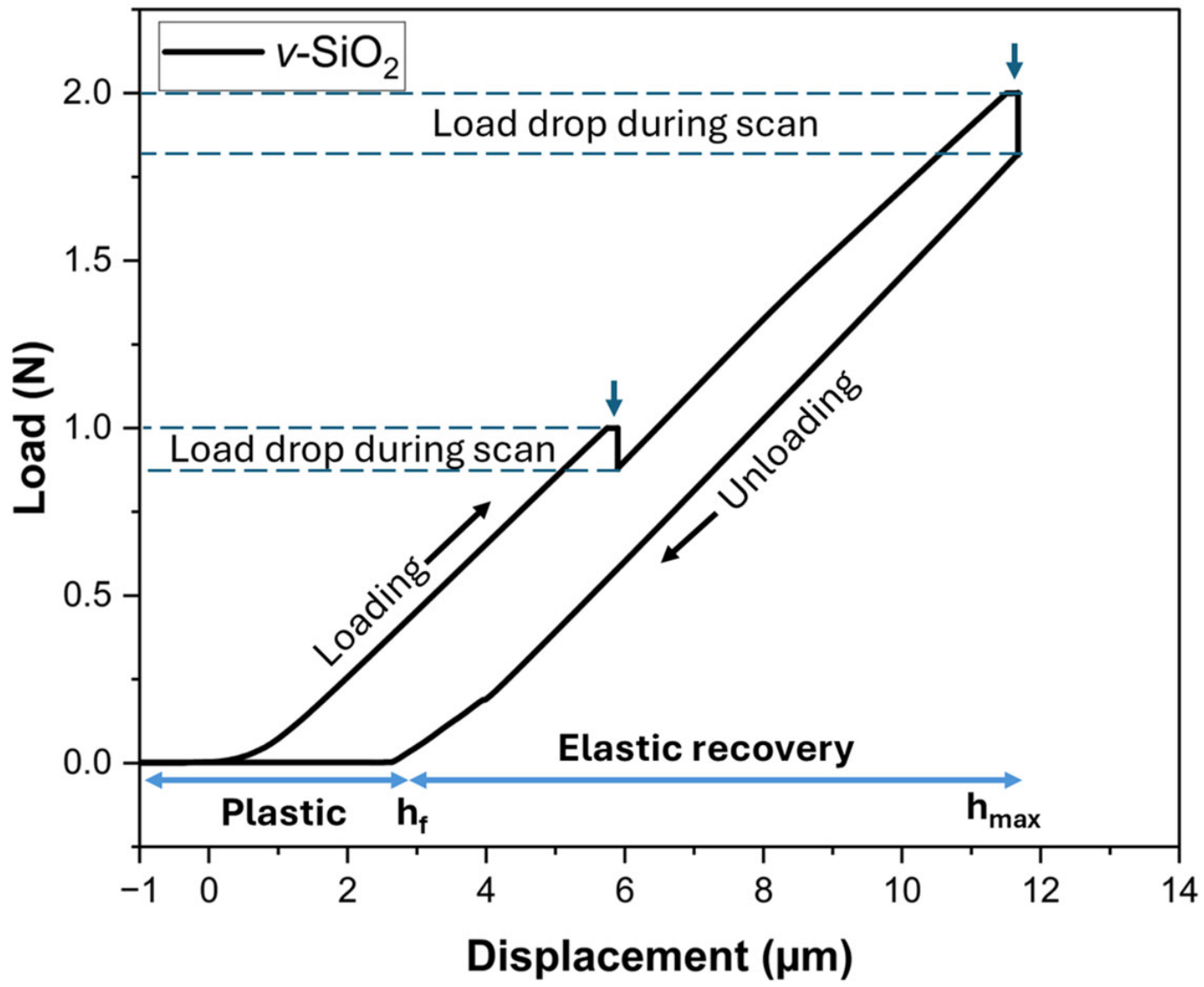

**Figure S7.** An example of load-displacement curve obtained during in situ indentation shown for vitreous silica. Displacement is kept constant at different load intervals for diffraction mapping. We note that load-drops, as highlighted with the arrows, were observed during the displacement hold segment of scanning due to creep.

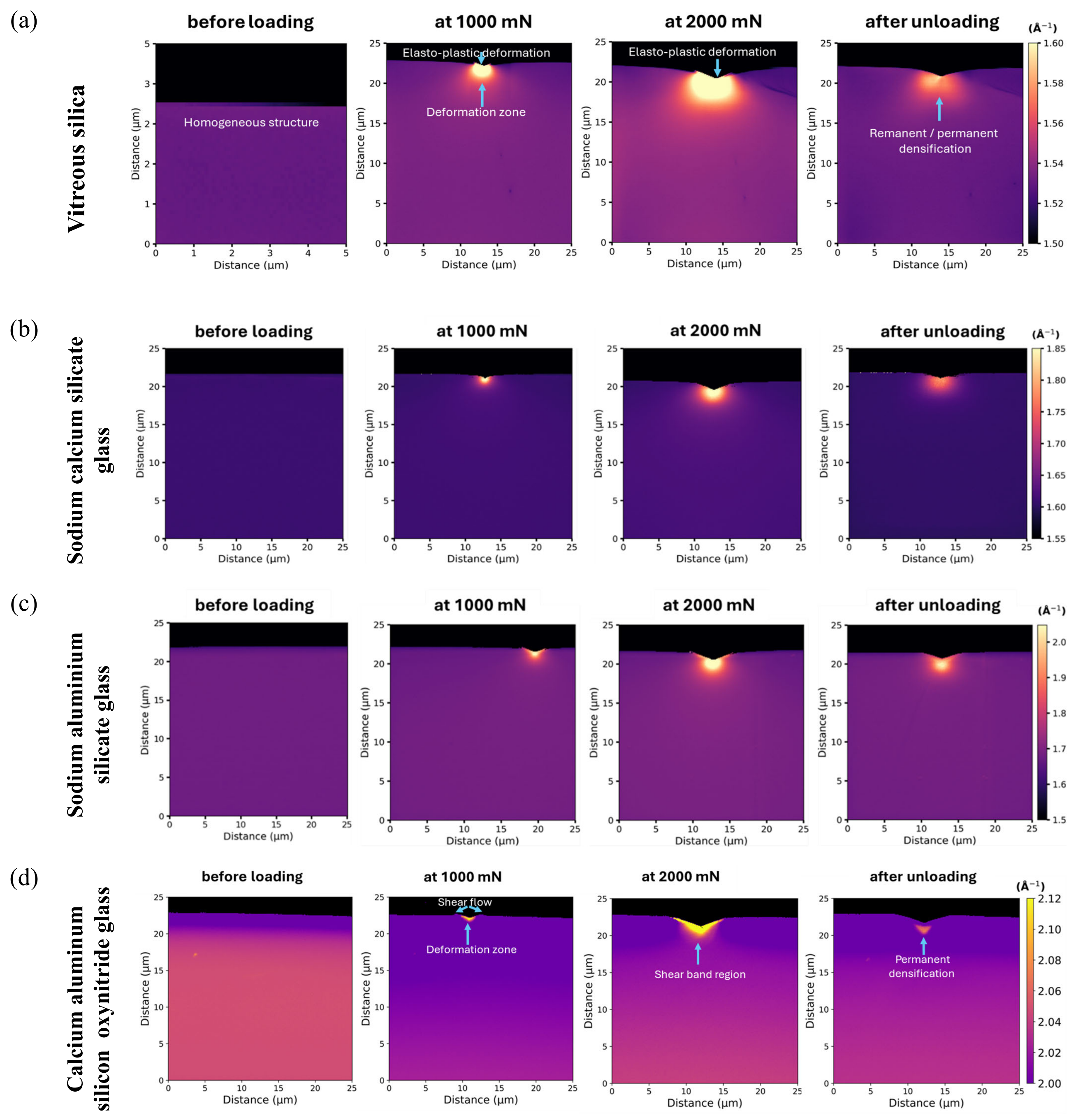

**Figure S8.** FSDP position maps obtained over different indentation loading steps on (a) vitreous silica, (b) NCS, (c) NAS, and (d) CAS-N glasses. Black region in the map corresponds to air regions.

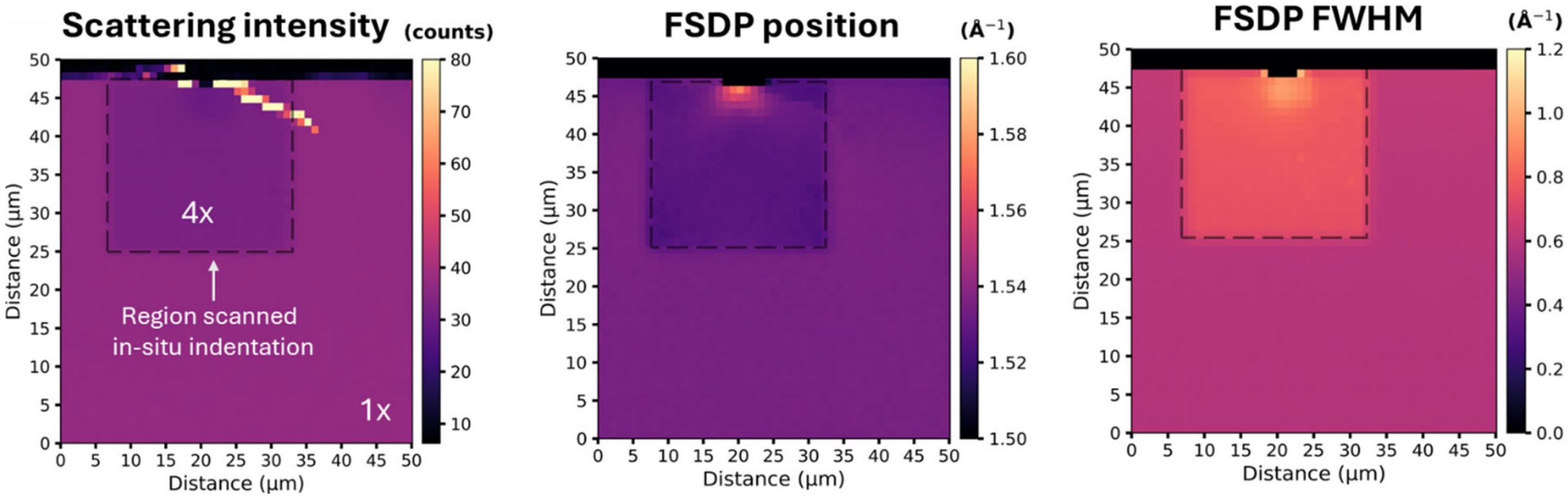

**Figure S9.** 2D maps of the change in intensity, position, and full width at half maximum of the FSDP between regions in the sample exposed with the X-ray beam for different number of times. Results are shown for vitreous silica.

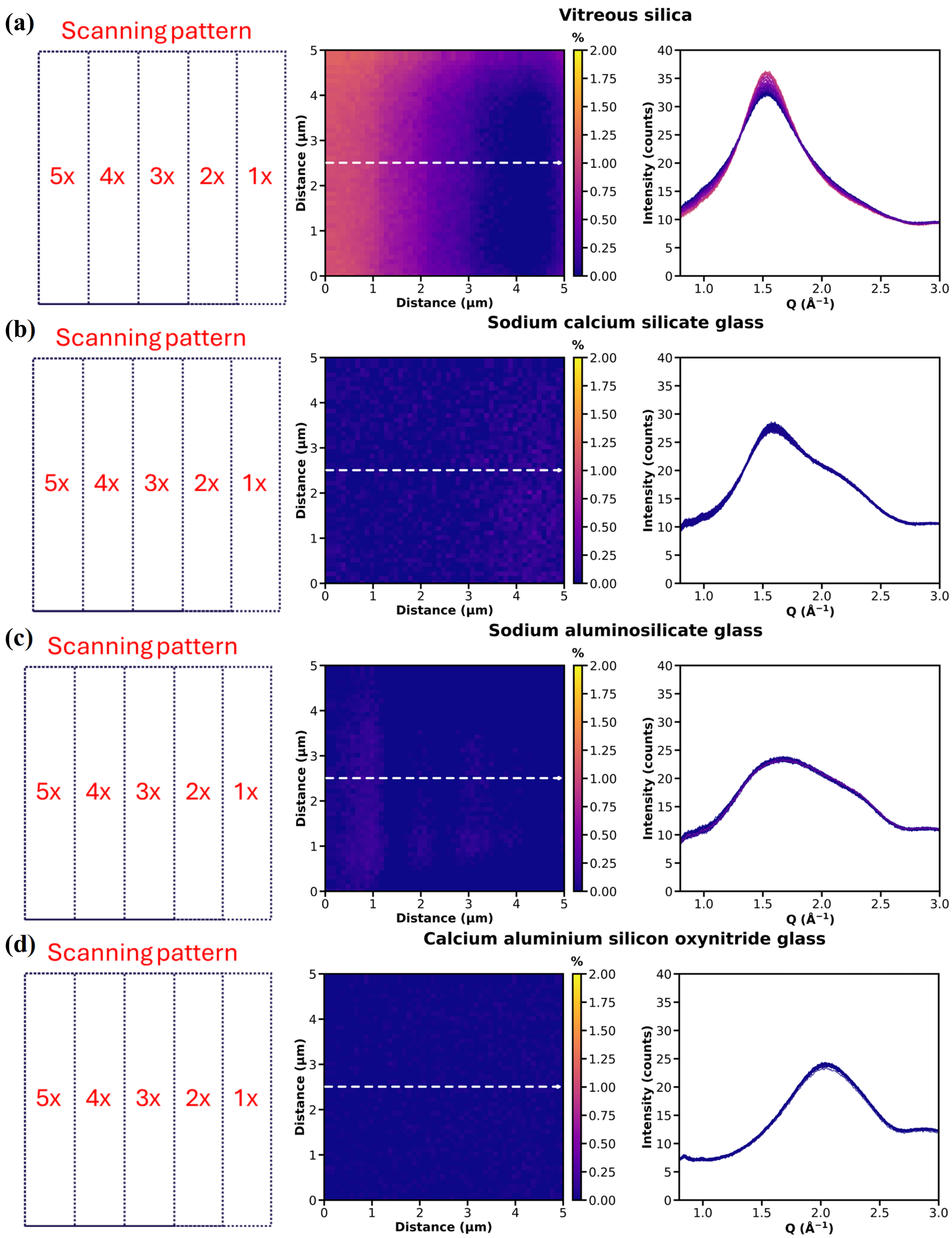

**Figure S10.** Schematic illustration of the multiple small area scans performed over a given region on the glass and FSDP position maps showing beam-induced structural changes (or lack thereof) for (a) vitreous silica, (b) sodium calcium silicate, (d) sodium aluminosilicate, and (d) oxynitride calcium aluminosilicate glasses.

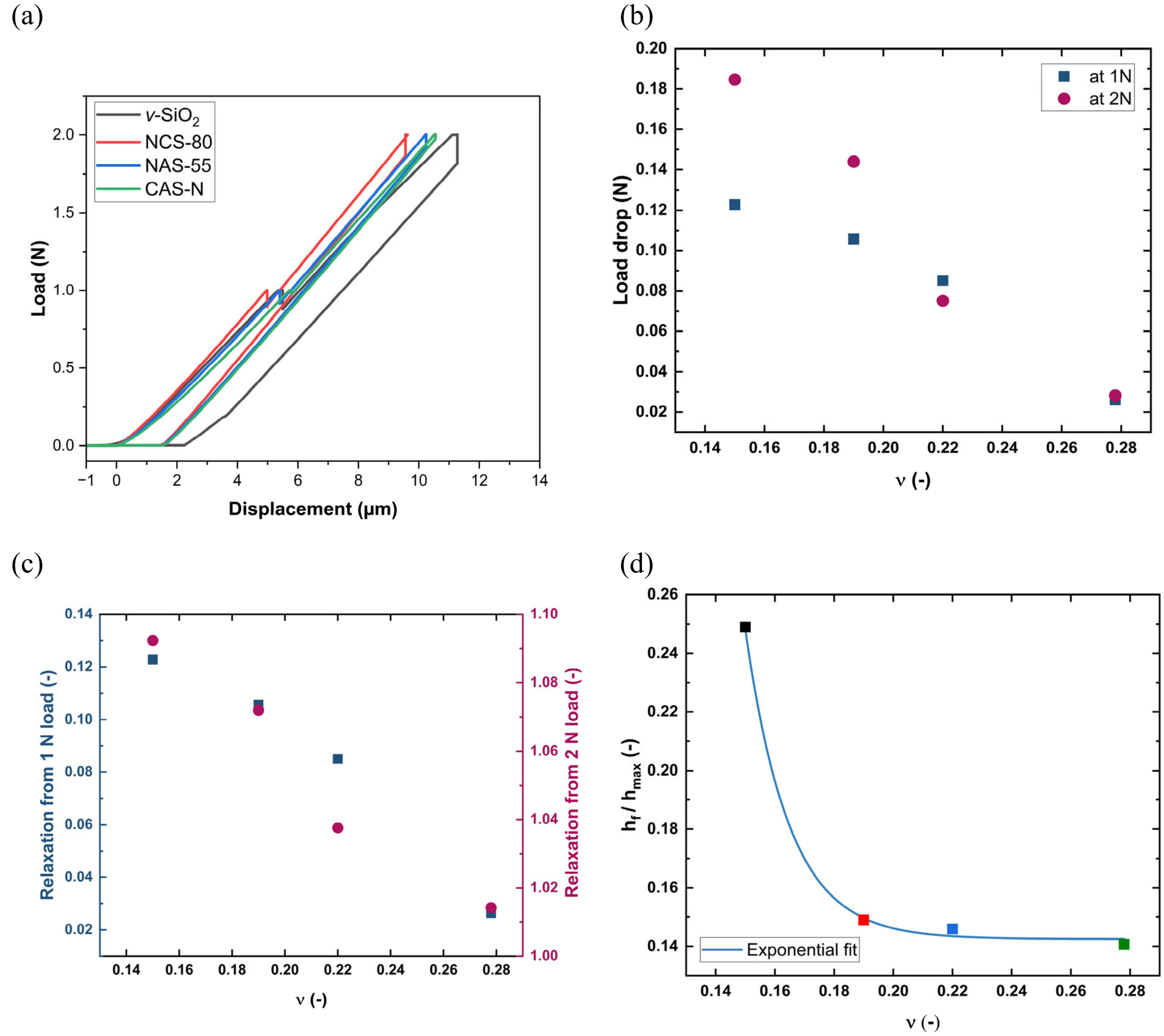

**Figure S11.** (a) Load-displacement curves obtained during indentation of four different glasses, which are all normalized to having the same starting displacement value of zero. (b) Relation between the extent of load drop during constant displacement (at 1 and 2 N load) and the Poisson's ratio of the studied glasses. (c) Dependence of the load relaxation (at 1 and 2 N load) on the Poisson's ratio of the studied glasses. (d) Dependence of the extent of permanent deformation left after complete unloading on the Poisson's ratio of the studied glasses.

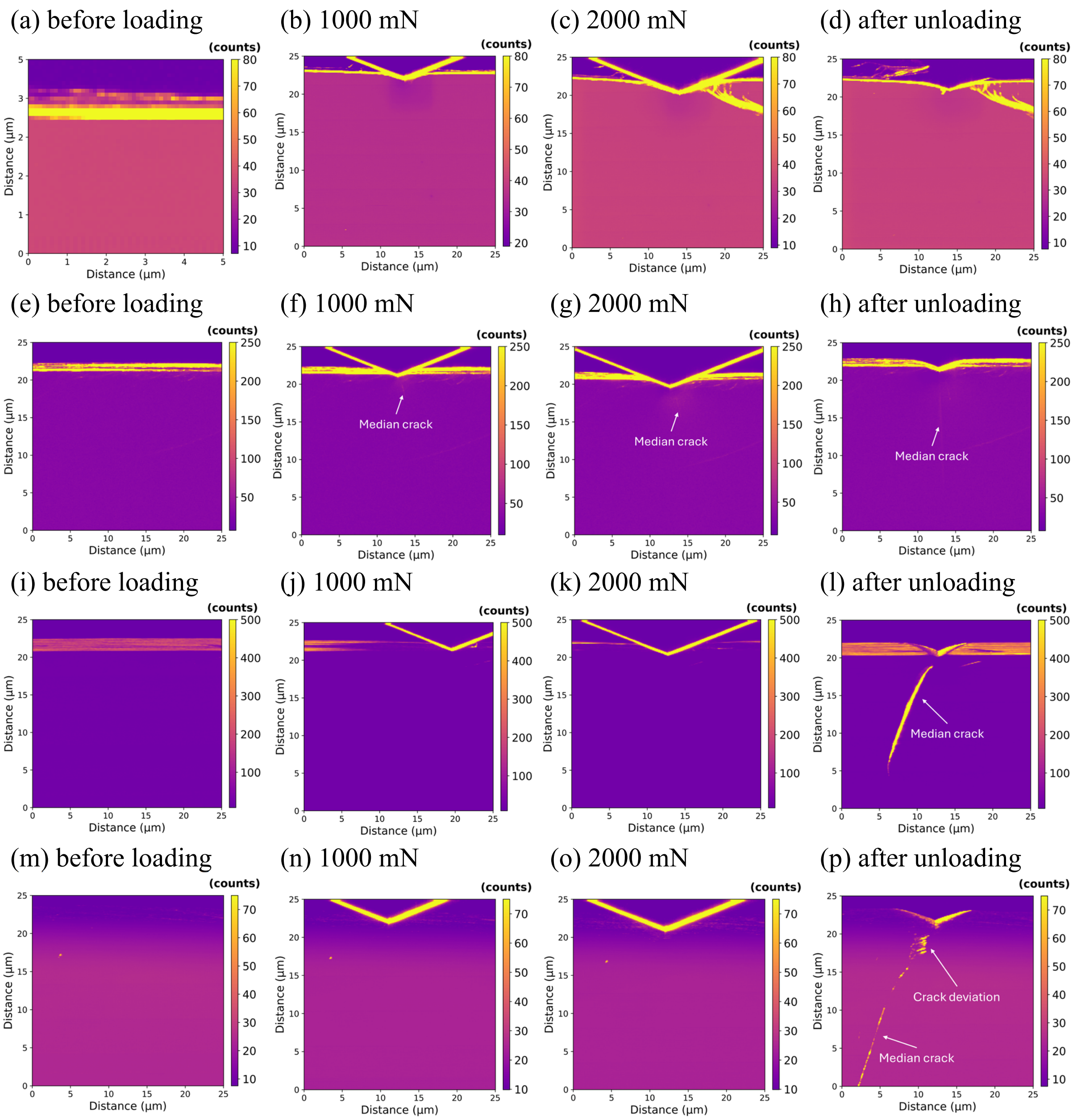

**Figure S12.** Mapping of the FSDP intensity obtained over different indentation loading steps from (a-d) vitreous silica, (e-h) NCS, (i-l) NAS, and (m-p) CAS-N glasses. Inverted triangle in each of the indentation loading scans correspond to the indenter tip, which along with the sample surface and defects scatter the X-rays much more than the bulk of the sample, resulting in higher scattering intensities.

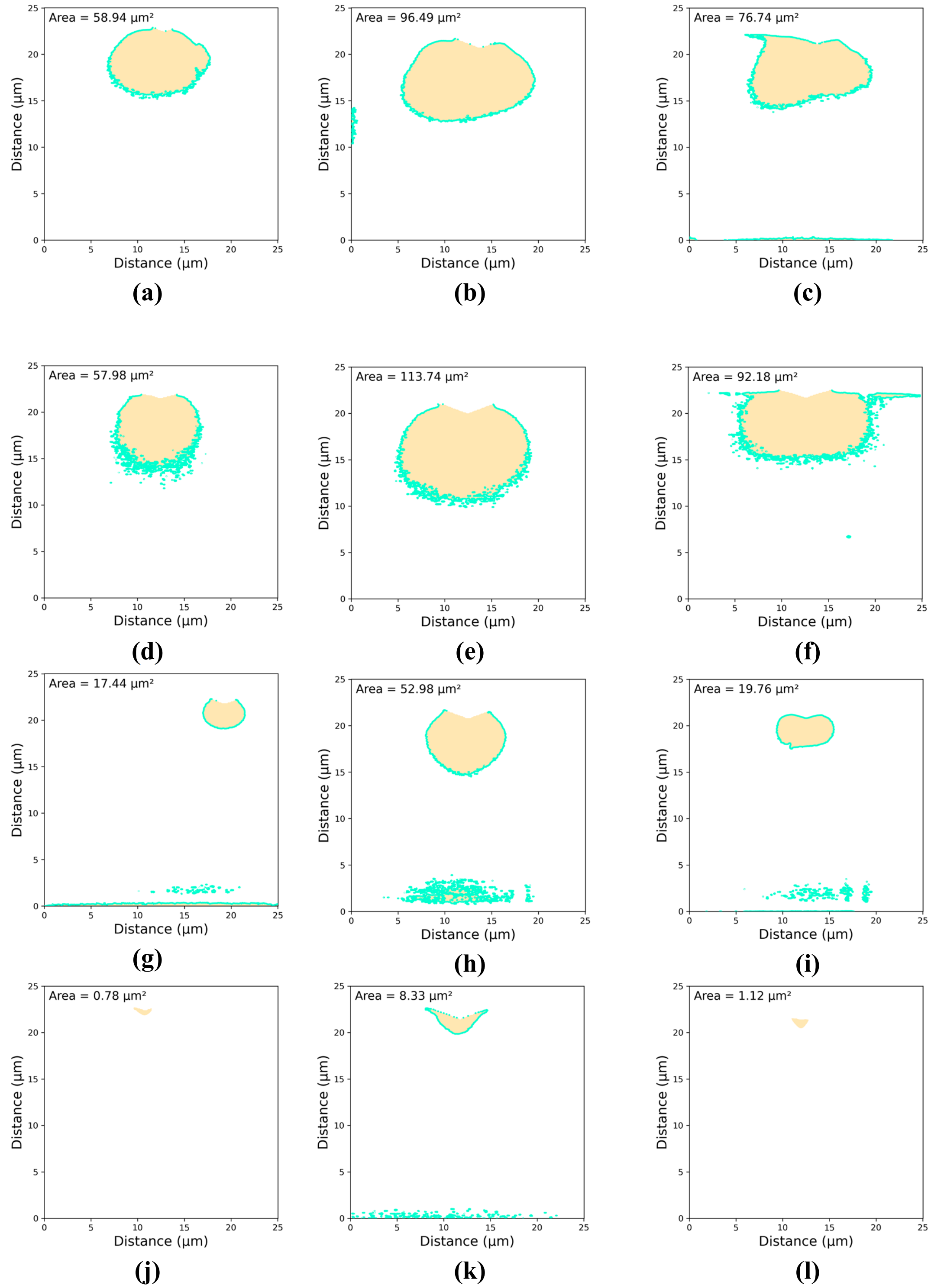

**Figure S13.** Segmentation of FSDP position maps (based on thresholding method) to obtain the deformation zones. Left column panels show glasses at the 1 N load, center column panels show show glasses at the maximum load (2 N), while right column panels show glasses after unloading. Results are shown for (a,b) vitreous silica, (c,d) NCS, (e,f) NAS, and (g,h) CAS-N glasses.